\documentclass[a4paper,12pt,autodetect-engine]{article}

\usepackage{amsmath}
\usepackage{amssymb}
\usepackage{amsfonts}
\usepackage{amsthm}
\usepackage{color}
\usepackage{cite}
\usepackage{slashed} 

\usepackage{ascmac}
\newcommand{\dop}{\mathrm{d}}
\newcommand{\Italicsf}[1]{\text{\itshape\sffamily #1}} 
\newcommand{\Isg}{\Italicsf{g}}
\newcommand{\IsB}{\Italicsf{B}}
\newcommand{\Isx}{\Italicsf{x}\hspace{1.5pt}}
\newcommand{\Isy}{\Italicsf{y}\hspace{1.5pt}}
\newcommand{\IsX}{\Italicsf{X}}
\newcommand{\IsH}{\Italicsf{H}}
\newcommand{\Isd}{\Italicsf{d}}
\newcommand{\Isphi}{\mathsf{\Phi}}
\newcommand{\IsGamma}{\mathsf{\Gamma}}

 \newcommand{\del}{\partial}

  \makeatletter
  \@addtoreset{equation}{section}
  \makeatother

\date{empty}
\begin{document}
\begin{titlepage}
\null
\begin{flushright}
October, 2020 \\
\texttt{2010.02701}
\end{flushright}
\vskip 2cm
\begin{center}
{\Large \bf 
World-volume Effective Theories \\
\vspace{0.5cm}
of Locally Non-geometric Branes 
}
\vskip 1.5cm
\normalsize
\renewcommand\thefootnote{\alph{footnote}}

{\large
Kenta Shiozawa\footnote{k.shiozawa(at)sci.kitasato-u.ac.jp}
and
Shin Sasaki\footnote{shin-s(at)kitasato-u.ac.jp}
}
\vskip 0.7cm
  {\it
  Department of Physics,  Kitasato University \\
  Sagamihara 252-0373, Japan
}
\vskip 2cm
\begin{abstract}
We study world-volume effective theories of 
five-branes in type II string theories.
We determine the bosonic zero-modes of the 
NS5-brane, the Kaluza-Klein monopole, the exotic Q5-, R5-branes and a
 space-filling brane, by direct calculations within the formalism of double field theory (DFT). 
We show that these zero-modes are Nambu-Goldstone modes associated with
 the spontaneously broken gauge symmetries in DFT.
They are organized into the bosonic part of the six-dimensional
 $\mathcal{N} = (1,1)$ vector and the $\mathcal{N} = (2,0)$ tensor multiplets.
Among other things, we examine the locally non-geometric R5-branes and space-filling branes 
that are characterized by the winding space. 
We also study effective theories of five-branes with string worldsheet instanton corrections.
\end{abstract}
\end{center}

\end{titlepage}

\newpage
\setcounter{footnote}{0}
\renewcommand\thefootnote{\arabic{footnote}}
\pagenumbering{arabic}
\tableofcontents

\section{Introduction} \label{sect:introduction}
U-duality is a key ingredient to understand whole structure of M-theory
and string theories \cite{Hull:1994ys}.
The duality exhibits its remarkable features in lower dimensions. 
For example, M-theory compactified on $T^D$ shows the U-duality group
$E_{D(D)}$ in the $(11-D)$-dimensional space-time.
BPS states in lower dimensions, 
which originate from the higher-dimensional branes that wrap $T^D$, are
classified by the U-duality group \cite{Blau:1997du, Obers:1998fb}.
In addition to the conventional branes such as the D-branes and the
NS5-branes and so on,
there appear yet unfamiliar objects in the
U-duality multiplets, known as the exotic branes \cite{Blau:1997du,
Obers:1998fb, Eyras:1999at, deBoer:2010ud}.
Detailed classifications of exotic branes are found, for example, in
\cite{Fernandez-Melgarejo:2018yxq, Berman:2018okd}.

The exotic branes are characterized by their highly non-perturbative
nature in string theory.
For example, they have tensions proportional to $g_s^{-\alpha}$ with $\alpha \ge 2$
which contrast to those of the D-branes given by the order
$g_s^{-1}$. Here $g_s$ is the string coupling constant.
Among other things, exotic branes appearing in the T-duality orbit of five-branes
in type II theories have been intensively studied \cite{deBoer:2010ud,
Kikuchi:2012za, Kimura:2013khz, Okada:2014wma, Sakatani:2014hba, Kimura:2014bea}.
There are five-branes denoted as $5^k_2$ \cite{Obers:1998fb} related by
T-duality transformations.
The ``2'' in the lowerscript represents that the tensions of the
five-branes are proportional to $g_s^{-2}$ 
while $k = 0,1,2,3,4$ stands for the number of the isometries in the
transverse directions in ten dimensions.
For example, $k = 0$ and $k=1$ correspond to 
the NS5-brane and the Kaluza-Klein (KK)-monopole.
The $k = 2$ case is an example of the exotic brane conventionally refered
as Q-brane \cite{Hassler:2013wsa}.
The geometry of the $5^2_2$-brane is patched together not by the diffeomorphism or gauge transformations 
but by T-duality transformations. 
In this sense, the background of the $5^2_2$-brane is known as a globally
non-geometric space or a T-fold \cite{Hull:2004in}.
Although it is a codimension two brane and its existence as a stand
alone object suffers from a subtle issue, 
one can explicitly write down the $5^2_2$-brane solution in the ordinary supergravity \cite{deBoer:2010ud}.
On the other hand, the $k=3$ brane has less obvious geometric meaning.
One can not perform a T-duality transformation in the $5^2_2$-brane since any further isometries are forbidden there.
However, from the viewpoint of the U-duality multiplet in lower dimensions, 
the $5^3_2$-brane should be allowed in higher dimensions.
Nevertheless, it is less obvious whether that is a solution to supergravity or not.

Double field theory (DFT) \cite{Hull:2009mi}, based on the doubled
formalism developed in \cite{Siegel:1993xq}, is an effective gravity
theory that has manifest T-duality in string theory.
Due to the $O(D,D)$ covariance, one can treat the T-duality orbit of
the branes without difficulty in DFT.
The action of DFT is given by a generalized Ricci scalar 
and it reduces to the action of supergravity with the imposition of
the strong constraint. 
Indeed, supergravity solutions such as the F-strings, the waves, the
NS5-branes, the KK-monopoles and even the globally non-geometric objects, are found to be solutions to DFT
\cite{Berkeley:2014nza, Berman:2014jsa}.
This is conceivable since DFT is, in a sense, a reformulation of supergravity.
However, the most remarkable difference between the ordinary supergravity
theories and DFT is the isometries for T-duality transformations.
Contrary to supergravities, DFT does not necessarily require isometries to perform the
T-duality transformation.
Therefore, one can perform the formal (non-isometric) T-duality
transformation to obtain geometries that are not solutions to supergravity.
The $5^3_2$-brane is one of such kind of solution in DFT
\cite{Bakhmatov:2016kfn,Kimura:2018hph}. 
Due to the absence of the isometries, the $5^3_2$-brane geometry
contains the dual coordinate $\tilde{x}$-dependence.
This $\tilde{x}$-dependence is no longer removed even by the T-duality
transformation and it is an essential nature of the new geometry.
This kind of brane is known as an R-brane or {\it the locally non-geometric
object} \cite{Hassler:2013wsa, Plauschinn:2018wbo, Berman:2020tqn}.
The geometry of the R-brane necessarily depends on the winding
coordinate $\tilde{x}$. Therefore it loses the ordinary geometric
meaning based on the point particle.
We can continue this procedure further.
The compactification on $T^4$ allows us to find the space-filling $5^4_2$-brane with
the non-trivial winding coordinate dependence.

The other place that appears the winding coordinate dependence of
geometries is the worldsheet instanton effects.
This is first proposed in \cite{Gregory:1997te} where the dual side of
the localized (unsmeared) NS5-brane is analyzed.
In \cite{Tong:2002rq}, it is shown that the isometry in the H-monopole
(the NS5-brane smeared along the transverse direction) is broken by the string worldsheet instanton effects. 
In the dual KK-monopole side, the worldsheet instanton effects induce the
dual coordinate dependence in the Taub-NUT geometry \cite{Harvey:2005ab,
Jensen:2011jna}. 
The modified geometry is known as the localized KK-monopole \cite{Berman:2014jsa}.
The same is true even for the globally non-geometric $5^2_2$-brane
\cite{Kimura:2013fda, Kimura:2013zva, Kimura:2018ain}. The geometry is modified and
characterized by the winding coordinate. We call it the localized Q5-brane.
They are also classified into the locally non-geometric objects.
A remarkable fact is that the localized KK-monopole and the localized Q5-brane
are shown to be solutions to DFT \cite{Kimura:2018hph}.
The geometry with the winding corrections are necessary to understand a
stringy nature of space-time and it is also discussed in other places
\cite{Lust:2017jox, Achmed-Zade:2018rfc}.

The purpose of this paper is to study the worldvolume effective theories of the locally
non-geometric five-branes. 
Among other things, we focus on the R5-brane ($5^3_2$-brane), the
space-filling $5^4_2$-brane, the localized KK-monopole, Q5-brane and also the
localized R5-brane, space-filling branes.
For the globally non-geometric objects, 
the world-volume effective actions of the Dirac-Born-Infeld (DBI) type 
are obtained through the duality covariantization or formal
T-duality transformations of known branes \cite{Chatzistavrakidis:2013jqa, Kimura:2014upa,
Kimura:2016anf, Sakatani:2016sko, Blair:2017hhy, Sakatani:2020umt}.
We here derive the effective theories of the locally non-geometric
five-branes by the direct calculations within the formalism of DFT.
We will explicitly determine the precise bosonic zero-modes of the five-branes
and show that they are Nambu-Goldstone modes associated with the broken
gauge symmetries in DFT. 
These zero-modes appear in the world-volume effective theories of the five-branes.
We will show that these zero-modes are organized into the bosonic
parts of the six-dimensional $\mathcal{N} = (1,1)$ vector and the $\mathcal{N}
= (2,0)$ tensor multiplets.

The organization of this paper is as follows.
In the next section, we briefly introduce five-brane solutions in DFT. 
In Section \ref{sct:zero-modes}, we determine the translational
zero-modes in the doubled space. 
In Section \ref{sct:tensor}, we determine the tensor zero-modes
associated with the spontaneous breaking of the RR gauge symmetries in DFT.
We will show that these zero-modes are organized into the desired
six-dimensional supermultiplets.
In Section \ref{sct:winding}, we focus on the world-volume effective
theories of the locally non-geometric five-branes.
Section \ref{sct:conclusion} is conclusion and discussions.
Appendix \ref{sct:DFT} devoted to a quick introduction to DFT.
Appendix \ref{sct:detailed_calculations} contains the detailed
calculations on the zero-modes.

\section{Locally non-geometric branes in double field theory} \label{sct:five-branes_DFT}
In this section, we give a brief introduction to locally non-geometric
branes, especially focusing on five-branes in DFT \cite{Kimura:2018hph}.
We consider a T-duality orbit of five-branes in type
II string theories and exhibit the locally non-geometric objects in DFT.

The dynamical fields in DFT of type II string theories are given by the generalized metric
$\mathcal{H}_{MN}$, the generalized dilaton $d$ and the $O(D,D)$
spinor $\chi$\footnote{
Unless otherwise stated, we consider $D=10$ in the following.
}.
The DFT action is given by \cite{Hohm:2010pp,Hohm:2011zr,Hohm:2011dv}:
\begin{align}
S_{\text{DFT}} (\mathbb{S}, d, \chi) = \int \! {\dop}^{2D} X \, 
\Bigg(
e^{-2d} \, \mathcal{R} (\mathcal{H}, d) + \frac{1}{4}
 (\slashed{\partial} \chi)^{\dagger} \mathbb{S} (\slashed{\partial} \chi)
\Bigg).
\label{eq:DFT_action}
\end{align}
The generalized Ricci scalar $\mathcal{R}$ is defined by 
\begin{align}
\mathcal{R} =& \
4 \mathcal{H}^{MN} \partial_M \partial_N d - \partial_M \partial_N
 \mathcal{H}^{MN} - 4 \mathcal{H}^{MN} \partial_M d \partial_N d
+ 4 \partial_M \mathcal{H}^{MN} \partial_N d
\notag \\
&
+ \frac{1}{8}
 \mathcal{H}^{MN} \partial_M \mathcal{H}^{KL} \partial_N
 \mathcal{H}_{KL}
 - \frac{1}{2} \mathcal{H}^{MN} \partial_M
 \mathcal{H}^{KL} \partial_K \mathcal{H}_{NL}.
\label{eq:generalized_Ricci_scalar}
\end{align}
where the indices $M,N, \ldots$ run over the $2D$ dimensions.
The inverse of the generalized metric $\mathcal{H}^{MN}$ is introduced
by $\mathcal{H}^{MP} \mathcal{H}_{PN} = \delta^M {}_N$ with its defining
$O(D,D)$ property $\mathcal{H}^{MN} = \eta^{MP} \eta^{NQ}
\mathcal{H}_{PQ}$.
Here $\eta^{MN}$ and its inverse $\eta_{MN}$ are $O(D,D)$ invariant
metric given by
\begin{align}
\eta_{MN} = 
\begin{pmatrix}
0 & \delta^{\mu} {}_{\nu}
\\
\delta_{\mu} {}^{\nu} & 0
\end{pmatrix}, \qquad 
\eta^{MN} = 
\begin{pmatrix}
0 & \delta_{\mu} {}^{\nu}
\\
\delta^{\mu} {}_{\nu} & 0
\end{pmatrix}.
\end{align}
The coordinate of the $2D$-dimensional doubled space is decomposed as $X^M = (\tilde{x}_{\mu},
x^{\mu}), \, (\mu = 0, \ldots, D-1)$ 
where $x^{\mu}$ are the Fourier conjugate of the KK modes (the geometric
coordinates) while $\tilde{x}_{\mu}$ are the Fourier conjugate of the
winding modes (the winding coordinates).
The $O(D,D)$ spinor $\chi$ is assumed to satisfy 
the self-duality constraint $\slashed{\partial} \chi = -
\mathcal{K} \slashed{\partial} \chi$.
Here $\slashed{\partial} = \Gamma^M \partial_M$ and $\Gamma^M$ are the
$2D$-dimensional gamma matrices.
The operator $\mathcal{K} = C^{-1} \mathbb{S}$ is defined by the charge
conjugation matrix $C$ and the spinor representation of the generalized
metric $\mathbb{S} = \mathbb{S}^{\dagger} \in \mathrm{Spin}^{-} (D,D)$.
See Appendix \ref{sct:DFT} and \cite{Hohm:2011dv} for the precise definitions.

The action \eqref{eq:DFT_action} makes the $O(D,D)$ T-duality be
manifest and it is invariant under the following gauge transformations:
\begin{align}
\delta_{\xi} \mathcal{H}_{MN} =& \ \widehat{\mathcal{L}}_{\xi} \mathcal{H}_{MN} = 
\xi^P \partial_P \mathcal{H}^{MN} + (\partial^M \xi_P - \partial_P
 \xi^M) \mathcal{H}^{PN} + (\partial^N \xi_P - \partial_P \xi^N) \mathcal{H}^{MP},
\notag \\
\delta_{\xi} d =& \ \widehat{\mathcal{L}}_{\xi} d = \xi^M \partial_M d -
 \frac{1}{2} \partial_M \xi^M,
\notag \\
\delta_{\xi} \chi =& \ \widehat{\mathcal{L}}_{\xi} \chi = \xi^M \partial_M \chi 
+ \frac{1}{2} \partial_N \xi_M \Gamma^N \Gamma^M \chi,
\notag \\
\delta_{\lambda} \chi =& \ \slashed{\partial} \lambda,
\label{eq:DFT_gauge_transformations}
\end{align}
where $\widehat{\mathcal{L}}_{\xi}$ is the generalized Lie derivative and
$\xi^M, \lambda$ are parameters of a doubled vector and an $O(D,D)$ spinor.
The gauge invariance of the DFT action \eqref{eq:DFT_action} is
guaranteed by the so-called strong constraint $\partial^M * \partial_M * = 0$.
Here $*$ stands for any dynamical fields and the gauge parameters in DFT.

It is convenient to parametrize the DFT fields in terms of the ordinary
supergravity fields:
\begin{align}
\mathcal{H}_{MN} = 
\begin{pmatrix}
g^{\mu \nu} & - g^{\mu \rho} B_{\rho \nu}
 \\
B_{\mu \rho} g^{\rho \nu} & g_{\mu \nu} - B_{\mu \rho} g^{\rho \sigma} B_{\sigma \nu}
\end{pmatrix},
\qquad 
e^{-2d} = \sqrt{-g} e^{-2\Phi}.
\label{eq:gm_parametrization}
\end{align} 
where $g_{\mu \nu}, B_{\mu \nu}, \Phi$ are identified with the space-time metric, the NSNS
$B$-field and the dilaton in type II supergravities.
Meanwhile, the $O(D,D)$ spinor is expanded as
\begin{align}
\chi = \sum_{p=0}^D \frac{1}{p!} C_{\mu_1 \cdots \mu_p} \psi^{\mu_1}
 \cdots \psi^{\mu_p} |0 \rangle,
\end{align}
where $C_{\mu_1 \cdots \mu_p}$ are identified with the RR $p$-forms%
\footnote{We follow the convention of the RR $p$-forms employed in \cite{Hohm:2011zr,Hohm:2011dv}.}.

The equations of motion in DFT are given by
\begin{align}
&
\mathcal{R}_{MN} + e^{2d} \mathcal{E}_{MN} = 0, 
\qquad 
\mathcal{R} = 0,
\qquad 
\slashed{\partial} (\mathcal{K} \slashed{\partial} \chi) = 0.
\label{eq:eom}
\end{align}
Here the generalized Ricci tensor $\mathcal{R}_{MN}$ and the energy-momentum
tensor $\mathcal{E}_{MN}$ are defined by
\begin{align}
\mathcal{R}_{MN} = P_{MN} {}^{KL} \mathcal{K}_{KL}, \qquad 
\mathcal{E}^{MN} = - \frac{1}{16} \mathcal{H}^{(M}
 \overline{\slashed{\partial} \chi} \Gamma^{N)P} \slashed{\partial} \chi.
\end{align}
The projection operator $P$ and the tensor $K$ are given by the
following expressions:
\begin{align}
P_{MN} {}^{KL} &=  
\frac{1}{2} \left[
\delta_M {}^{(K} \delta_N {}^{L)} - \mathcal{H}_{MP} \eta^{P (K}
 \eta_{NQ} \mathcal{H}^{L) Q}
\right],
\label{eq:projection}
\\
{\mathcal K}_{MN} &=
	{1 \over 8} \partial_M {\mathcal H}^{KL} \partial_N {\mathcal H}_{KL}
	- {1 \over 4} (\partial_L - 2\partial_L d) {\mathcal H}^{KL} \partial_K {\mathcal H}_{MN}
	+ 2 \partial_M \partial_N d
\notag \\
& \quad
	- {1 \over 2} \partial_{(M} {\mathcal H}^{KL} \partial_L {\mathcal H}_{N)K}
	+ {1 \over 2} (\partial_L - 2\partial_L d) \left[ {\mathcal H}^{KL} \partial_{(M} {\mathcal H}_{N)K}
		+ \eta^{KP} \eta^{LQ} {\mathcal H}_{P(M} \partial_K
 {\mathcal H}_{N)Q} \right].
\label{eq:K_tensor}
\end{align}
We use the convention of the (anti)symmetrization with the weight factor
such as $A_{(M} B_{N)} = (A_M B_N + A_N B_M)/2$.

Now we introduce the five-brane solutions in DFT. 
We consider the ansatz of the localized DFT monopole
\cite{Berman:2014jsa}: 
 \begin{align}
 {\IsH}_{MN}
 &= 
 \begin{pmatrix}
 \eta^{mn} & 0 & 0 & 0 \\
 0 & H^{-1} \delta^{ab} & 0 & - H^{-1} b^a{}_b \\
 0 & 0 & \eta_{mn} & 0 \\
 0 & H^{-1} b_a{}^b & 0 & H \delta_{ab}
 \end{pmatrix}, &
 {\Isd}
 &= \text{const.} - \frac{1}{2} \log H, &
 \chi 
 &=  0.
 \label{eq:DFT_monopole_ansatz}
 \end{align}
We introduce a new decomposition of the doubled coordinate:
\begin{align}
{\IsX}^M = (\bar{\Isx}_m, \bar{\Isy}_a; {\Isx}^m, {\Isy}^a),
\label{eq:ISx_coordinates}
\end{align}
where ${\Isx}^m \, (m = 0,1, \ldots, 5)$, ${\Isy}^a \, (a = 6, \ldots, 9)$
 are the worldvolume and the transverse coordinates and 
 $\bar{\Isx}_m, \bar{\Isy}_a$ are their duals.
We here stress that we never give the physical meaning of the coordinates
$({\Isx}^m, {\Isy}^a, \bar{\Isx}_m, \bar{\Isy}_a)$ at this stage.
Namely, the roles of the geometric and the winding coordinates are fixed
through the following $O(D,D)$ assignment:
\begin{align}
X^M &= \Omega^M{}_N {\IsX}^N, \qquad \Omega \in O(D,D).
\label{eq:ODD_assign}
\end{align}
One should be careful about the distinction between $X^M =
(\tilde{x}_{\mu}, x^{\mu})$ and ${\IsX}^M$ in
\eqref{eq:ISx_coordinates}.
Likewise, the generalized metric \eqref{eq:gm_parametrization} 
parametrized in the space-time metric $g$ and 
the NSNS $B$-field are read off by the ansatz
\eqref{eq:DFT_monopole_ansatz} through the corresponding $O(D,D)$ assignment:
\begin{align}
{\mathcal H}_{MN} &= {\IsH}_{KL} \Omega^K{}_M \Omega^L{}_N, 
\qquad \Omega \in O(D,D).
\label{eq:gm_duality_frame_fixing}
\end{align} 

For the function $H$ appearing in the ansatz
\eqref{eq:DFT_monopole_ansatz}, we assume that it is a function of the transverse coordinate ${\Isy}^a$.
Together with a trivial condition $\bar{\partial}^a * = 0$ to the strong
constraint, one can show that the ansatz \eqref{eq:DFT_monopole_ansatz} satisfies
the equations motion \eqref{eq:eom} when the following conditions are satisfied \cite{Kimura:2018hph}:
\begin{align}
3 \partial_{[a} b_{bc]} &= \varepsilon_{abcd} \, \partial^d H({\Isy}), \qquad
\Box H = 0.
\label{eq:BPS_condition}
\end{align}
Here $b_{ab}$ is an antisymmetric tensor depending on ${\Isy}^a$, 
$\varepsilon_{abcd}$ is the Levi-Civita symbol,
and $\Box = \delta^{ab} \frac{\del}{\del {\Isy}^a} \frac{\del}{\del {\Isy}^b}$ is the
flat Laplacian defined by ${\Isy}^a$.
Therefore $H$ is a harmonic function in the ${\Isy}$-space.
The equations \eqref{eq:BPS_condition} correspond to the BPS condition
for the NS5-brane in ordinary supergravity.
We stress that the condition \eqref{eq:BPS_condition} allows for five-branes
of any codimensions, namely, 
$H$ is a harmonic function in 0 to 4 dimensions.

In the following, we will fix specific T-duality frames by giving the
$O(D,D)$ matrices $\Omega$ explicitly and write down the five-brane solutions in each frame.

\paragraph{NS5-brane ($5^0_2$)}
When the $O(D,D)$ matrix is given by $\Omega = 1$, 
the role of the geometric coordinates is assigned to $x^\mu = ({\Isx}^m,
{\Isy}^a)$. 
The role of the winding coordinates is assigned to $\tilde{x}_\mu = (\bar{\Isx}_m, \bar{\Isy}_a)$.
According to the relation \eqref{eq:gm_duality_frame_fixing}, 
the metric, the $B$-field and the dilaton are read off  
\begin{align}
{\dop}s^2
&=
\eta_{mn} {\dop}{\Isx}^m {\dop}{\Isx}^n 
+ H \delta_{ab} {\dop}{\Isy}^a {\dop}{\Isy}^b ,
\notag \\
B
&= {1 \over 2} b_{ab} {\dop}{\Isy}^a \wedge {\dop}{\Isy}^b,
\notag \\
e^{2\Phi}
&= H.
\label{eq:NS5-brane}
\end{align}
This is nothing but the NS5-brane solution in type II supergravities.
It is obvious that the world-volume of the NS5-brane extends to the ${\Isx}^0, \cdots, {\Isx}^5$ directions.
The harmonic function $H(r)$ is defined in the transverse directions
${\Isy}^6, \cdots, {\Isy}^9$ and is given by
\begin{align}
H (r) = c_1 + \frac{c_2}{r^2}, \qquad r^2 = \delta_{ab} {\Isy}^a {\Isy}^b.
\end{align}
Here $c_1, c_2$ are appropriate constants.
Note that the explicit form of $b_{ab}$ is determined through the BPS condition \eqref{eq:BPS_condition}. 

\paragraph{KK-monopole ($5^1_2$)}
The KK-monopole is obtained by the T-duality transformation of the
NS5-brane along a transverse direction.
Remember that in order to perform the genuine T-duality transformations, we introduce isometries
along the transverse directions. 
To do so, we smear the harmonic function along the ${\Isy}^9$-direction in the NS5-brane.
Then $H$ becomes 
\begin{align}
& 
H (r) = c_1 + \frac{c_2}{r}, \quad 
r^2 = \delta_{ij} {\Isy}^i {\Isy}^j,
\end{align}
where $c_1,c_2$ are again appropriate constants.

We now give another $O(D,D)$ matrix $\Omega = h_9$, 
where $h_{\sf k}$ is a factorized T-duality transformation along the
${\Isy}^{\sf k}$-direction given by
\begin{align}
h_{\sf k}
&= 
\begin{pmatrix}
1 - e_{\sf k} & e_{\sf k} \\
e_{\sf k} & 1 - e_{\sf k}
\end{pmatrix}, \qquad
(e_{\sf k})_{\mu\nu} = \delta_{\mu {\sf k}} \delta_{\nu {\sf k}}. 
\end{align}
Then the assignment of the coordinates are $x^{\mu} = ({\Isx}^m,
 {\Isy}^i, \bar{\Isy}_9), \, (i = 6,7,8)$
and $\tilde{x}_{\mu} = (\bar{\Isx}_m, \bar{\Isy}_i, {\Isy}^9)$.
The BPS equation \eqref{eq:BPS_condition} is solved by a Dirac monopole 
$B_{i9} = A_i = b_{i9}$ in the $({\Isy}^6,{\Isy}^7,{\Isy}^8)$-plane.
The other components of the $B$-field are all zero.
Then we write down the explicit solution as 
\begin{align}
{\dop}s^2
&=
\eta_{mn} \, {\dop}{\Isx}^m {\dop}{\Isx}^n
+ H^{-1} ({\dop}\bar{\Isy}_9 + A_i {\dop}{\Isy}^i)^2 
+ H \delta_{ij} \, {\dop}{\Isy}^i {\dop}{\Isy}^j,
\notag \\
B
&= 	0,
\notag \\
e^{2\Phi} &= \text{const.}
\label{eq:KK-monopole}
\end{align}
This is nothing but the KK-monopole of codimension three in type II
supergravities.
The transverse direction is characterized by the Taub-NUT space whose
isometry direction is given by $\bar{\Isy}_9$.
The solution \eqref{eq:KK-monopole} is apparently geometric.

\paragraph{Q5-brane ($5^2_2$)}
The so-called exotic Q5-brane or $5^2_2$-brane is obtained by 
performing the T-duality transformation along the other direction in the KK-monopole.
We now give yet another $O(D,D)$ matrix $\Omega = h_8 \cdot h_9$. 
Then the geometric coordinates are $x^{\mu} = ({\Isx}^m, {\Isy}^{\alpha}, \bar{\Isy}_8, \bar{\Isy}_9) \,
(\alpha = 6,7)$ while the winding ones are $\tilde{x}_{\mu} = (\bar{\Isx}_m,
\bar{\Isy}_{\alpha}, {\Isy}^8, {\Isy}^9)$. 
The new isometry direction is given by $\bar{\Isy}_8$.
The harmonic function becomes 
\begin{align}
H = c_1 + c_2 \log \frac{\mu}{r}, \quad 
r^2 = \delta_{\alpha \beta} {\Isy}^{\alpha} {\Isy}^{\beta},
\end{align}
where $c_1,c_2$ are constants and $\mu$ is a parameter that characterizes the effective description
of defect branes as a stand alone object \cite{Bergshoeff:2011se, Kikuchi:2012za}.
Then the resulting solution is given by 
\begin{align}
{\dop}s^2
&= \eta_{mn} \, {\dop}{\Isx}^m {\dop}{\Isx}^n
	+ H \delta_{\alpha\beta} \, {\dop}{\Isy}^\alpha {\dop}{\Isy}^\beta
	+ {H \over H^2 + A_8^2} \big[ {\dop}\bar{\Isy}_9^2 + {\dop}\bar{\Isy}_8^2 \big],
\notag \\
B
&= - \frac{A_8}{H^2 + A_8^2} \, {\dop}\bar{\Isy}_8 \wedge {\dop}\bar{\Isy}_9,
\notag \\
e^{2\Phi} &= \frac{H}{H^2 + A_8^2}.
\label{eq:Q5-brane}
\end{align}
Here 
\begin{align}
A_8 = - c_2 \, \text{arctan} \left( \frac{{\Isy}^7}{{\Isy}^6} \right)
\end{align}
is the smeared Dirac monopole in a specific gauge.
The solution \eqref{eq:Q5-brane} has a non-trivial $O(2,2)$ monodromy around the
brane core and the geometry is not patched together by the diffeomorphism or the $B$-field gauge
transformation. Therefore the solution is not geometric in the usual sense and it is called a
globally non-geometric background or a T-fold \cite{Hull:2004in}.
However, it is obvious that the solution \eqref{eq:Q5-brane} is written
in the geometric coordinate and is indeed a solution to supergravity \cite{deBoer:2010ud}.

\paragraph{R5-brane ($5^3_2$)}
We now try to make further T-duality transformation along another
transverse direction. 
To this end, we give $\Omega = h_7 \cdot h_8 \cdot h_9$. 
Then $x^{\mu} = ({\Isx}^m, {\Isy}^6, \bar{\Isy}_{\hat{m}}), \, (\hat{m}
= 7,8,9)$ is the geometric and $\tilde{x}_{\mu} = (\bar{\Isx}_m, \bar{\Isy}_6, {\Isy}^{\hat{m}})$ are the winding coordinates.
The harmonic function becomes 
\begin{align}
H = c_1 + c_2 |{\Isy}^6|.
\end{align}
Here $c_1,c_2$ are constants again.
Then the 
solution reads
\begin{align}
{\dop}s^2
&= \eta_{mn} \, {\dop}{\Isx}^m {\dop}{\Isx}^n
+ H ({\dop}{\Isy}^6)^2
	+ H^{-1} {\dop}\bar{\Isy}_7^2
	+ \frac{H}{H^2 + A_8^2} \big[ {\dop}\bar{\Isy}_8^2 + {\dop}\bar{\Isy}_9^2 \big],
\notag \\
B
&= - \frac{A_8}{H^2 + A_8^2} \, {\dop}\bar{\Isy}_8 \wedge {\dop}\bar{\Isy}_9,
\notag \\
e^{2\Phi} &= \frac{1}{H^2 + A_8^2},
\label{eq:R5-brane}
\end{align}
where the smeared Dirac monopole becomes
\begin{align}
A_8 = c_2 (\mathrm{sgn} \, {\Isy}^6) {\Isy}^7.
\end{align}
Due to the BPS condition \eqref{eq:BPS_condition}, the $B$-field should
depend on 
${\Isy}^7$.
A remarkable feature is that the solution depends on ${\Isy}^7$ which is the
winding coordinate.
The solution \eqref{eq:R5-brane} is known as the R5-brane \cite{Hassler:2013wsa}.
Since it is impossible to represent the solution only by the geometric  
coordinates but it should include the dual winding coordinates $\tilde{x}_{7} = {\Isy}^7$,
this is called a locally non-geometric solution.
We stress that this is not a solution to conventional supergravity anymore but must be in string theory.
However, it is discussed that locally non-geometric branes of domain
wall type are solutions to deformed supergravities \cite{Fernandez-Melgarejo:2018yxq}.

\paragraph{SF5-brane ($5^4_2$)}
Finally, we consider a space-filling (SF) 5-brane \cite{Fernandez-Melgarejo:2018yxq, Kimura:2018hph}.
This is obtained by performing the T-duality transformation along
the final transverse direction. 
Given $\Omega = h_6 \cdot h_7 \cdot h_8 \cdot h_9$, 
the geometric coordinates are $x^{\mu} = ({\Isx}^m, \bar{\Isy}_a) \, (a =
6,7,8,9)$ while the winding ones are $\tilde{x}_{\mu} = (\bar{\Isx}_m, {\Isy}^a)$.
The solution becomes 
\begin{align}
{\dop}s^2
&= \eta_{mn} \, {\dop}{\Isx}^m {\dop}{\Isx}^n
+ H^{-1} \delta^{ab} \, {\dop}\bar{\Isy}_a {\dop}\bar{\Isy}_b
	- \frac{H^{-1}}{H^2 + A_8^2}
\big[ A_8^2 \, {\dop}\bar{\Isy}_8^2 + A_8^2 \, {\dop}\bar{\Isy}_9^2 \big],
\notag \\
B
&= - \frac{A_8}{H^2 + A_8^2} \, {\dop}\bar{\Isy}_8 \wedge {\dop}\bar{\Isy}_9,
\notag \\
e^{2\Phi} &= \frac{H^{-1}}{H^2 + A_8^2}.
\label{eq:sf-brane}
\end{align}
Here the harmonic function and the gauge field are given by
\begin{align}
H = c_1 + c_2 |{\Isy}^6|, \qquad 
A_8 = \frac{c_2}{2} (\mathrm{sgn} \, {\Isy}^6) {\Isy}^7.
\end{align}
We stress that ${\Isy}^6, {\Isy}^7$ are the winding coordinates and the solution
\eqref{eq:sf-brane} loses the nature of the conventional geometry.
This again shows the locally non-geometric property.
We note that if we introduce the complete isometries along the
four transverse directions by the smearing, the geometry of the space-filling brane
becomes trivial, namely, $H$ and $A_8$ are constants and no non-trivial
structure would be left.

\section{Translational zero-modes in doubled space} \label{sct:zero-modes} 
In this section, we determine the translational zero-modes of the five-branes
in the doubled space.
As for the ordinary branes in supergravity, the five-brane solutions
discussed in the previous section break the translational symmetries along the
transverse directions to the brane world-volume.
We remark that for the five-branes 
in the previous section, the transverse directions are realized by the $2 \times 4 = 8$ dimensional doubled
space.
The broken symmetries of the solutions give rise to the Nambu-Goldstone modes in DFT.
The translational symmetry in the doubled space is represented by the
generalized Lie derivative 
which is a part of the
gauge symmetry in \eqref{eq:DFT_gauge_transformations}. 
Once a solution $(\mathcal{H}_{MN}^{(0)}, d^{(0)}, \chi^{(0)})$ to DFT is found, we consider a fluctuation of fields
around the solution:
\begin{align}
\mathcal{H}_{MN}^{(0)} \to \mathcal{H}_{MN}^{(0)} + \delta
 \mathcal{H}_{MN},
\qquad 
d^{(0)} \to d^{(0)} + \delta d,
\qquad 
\chi^{(0)} \to \chi^{(0)} + \delta \chi
\label{eq:fluctuations}
\end{align}
where the variation $\delta$ is generated by 
broken symmetries.
We then substitute the fluctuations \eqref{eq:fluctuations}
into the equations of motion \eqref{eq:eom} and determine the zero-modes of the solution.

Given the fluctuations \eqref{eq:fluctuations}
around the five-brane
solution \eqref{eq:DFT_monopole_ansatz}, 
the variations of the equations of motion 
generated by the generalized Lie derivative $\widehat{\mathcal L}_\xi$  
become
\begin{align}
\delta \mathcal{R}_{MN} = 0,
\qquad 
\delta \mathcal{R} = 0.
\label{eq:NSNS_variations}
\end{align}
Note that since the $O(D,D)$ spinor $\chi$ all vanishes in the ansatz
\eqref{eq:DFT_monopole_ansatz} and it never breaks the translational
symmetry, the energy-momentum tensor $\mathcal{E}^{MN}$ does not contribute to
the equations \eqref{eq:NSNS_variations}.
In order to evaluate the variations in \eqref{eq:NSNS_variations}, we
decompose the generalized Ricci scalar and tensor into
the component form by using 
the following parametrization of the ansatz \eqref{eq:DFT_monopole_ansatz}:
\begin{align}
{\IsH}_{MN}
&= 
\begin{pmatrix}
{\Isg}^{\mu\nu} & - {\Isg}^{\mu\rho} {\IsB}_{\rho\nu} \\
{\IsB}_{\mu\rho} {\Isg}^{\rho\nu} 
& {\Isg}_{\mu\nu} - {\IsB}_{\mu\rho} {\Isg}^{\rho\sigma} {\IsB}_{\sigma\nu}
\end{pmatrix}, \quad
{\Isg}_{\mu\nu} = 
\begin{pmatrix}
\eta_{mn} & 0 \\
0 & H ({\Isx}) \delta_{ab}
\end{pmatrix}, \quad
{\IsB}_{\mu\nu} = 
\begin{pmatrix}
0 & 0 \\
0 & b_{ab} ({\Isx})
\end{pmatrix}, 
\notag \\
{\Isd}
&= {\Isphi} - {1 \over 4} \log |{\Isg}|, \quad
{\Isphi}
= \text{const.} + {1 \over 2} \log H ({\Isx}),
\label{eq:monopole(3.3)}
\end{align}
where ${\Isg}$, ${\IsB}$ are $D \times D$ symmetric and skew-symmetric
matrices and ${\Isphi}$ is a scalar function.
One should again keep in mind that their roles are not fixed at this
stage.
These matrices and scalar are identified with the space-time metric
$g_{\mu \nu}$, the $B$-field $B_{\mu \nu}$ and the dilation $\Phi$
through the $O(D,D)$ assignment \eqref{eq:ODD_assign} and \eqref{eq:gm_duality_frame_fixing}.
It is worthwhile to note that, with the imposition of the strong
constraint, the generalized Lie derivative
$\widehat{\mathcal{L}}_{\xi}$ by $\xi^M = (\bar{\epsilon}_\mu,
\epsilon^\mu)$ on the generalized metric and the dilaton 
\eqref{eq:monopole(3.3)} results in the linear combination of the 
ordinary Lie derivative $\mathcal{L}_{\epsilon}$ and the gauge
transformation $\delta_{\bar{\epsilon}}$ in their components.
Namely, we have  
\begin{align}
\delta {\IsH}_{MN} = \widehat{\mathcal L}_\xi {\IsH}_{MN}
& \xrightarrow{\bar{\partial}^\mu * = 0}
\left\{ 
\begin{aligned}
\delta {\Isg}_{\mu\nu} &= {\mathcal L}_\epsilon {\Isg}_{\mu\nu} \\
\delta {\IsB}_{\mu\nu} 
&= {\mathcal L}_\epsilon {\IsB}_{\mu\nu} + 2 \partial_{[\mu} \bar{\epsilon}_{\nu]}
\end{aligned}
\right. 
\notag \\
\delta {\Isd} = \widehat{\mathcal L}_\xi {\Isd}
& \xrightarrow{\bar{\partial}^\mu * = 0}
\delta {\Isphi} = {\mathcal L}_\epsilon {\Isphi}.
\end{align}

Before writing down the explicit variations of the components, we note the fact that the precise
Nambu-Goldstone modes in a curved space are given by the constant moduli, which will be promoted to the world-volume fields,
supplemented by appropriate weight
factors \cite{Adawi:1998ta}. 
Following the analysis in the DFT wave solution~\cite{Berkeley:2014nza},
we first try the ansatz for the gauge parameters $\xi^M =
(\bar{\epsilon}_\mu, \epsilon^\mu)$ given by 
\begin{align}
\epsilon^a = H^s \phi_0^a, \qquad
\bar{\epsilon}_a = H^s \bar{\phi}_{0a}, \qquad
(a = 6,7,8,9),
\label{eq:naive_ansatz}
\end{align}
where $\phi^a_0, \bar{\phi}_{0a}$ are the constant moduli parameters,
$H$ is the harmonic function in the solution
\eqref{eq:BPS_condition} and $s$ is a constant. 
However, one easily finds that the naive ansatz \eqref{eq:naive_ansatz}
allows only constant $\phi_0^a, \bar{\phi}_{0a}$ and it is not suitable
for the world-volume fields.
Instead, we mimic the ansatz 
discussed in the analysis on the M5-brane~\cite{Adawi:1998ta}:
\begin{align}
\epsilon^a = H^s \phi^a_0, \qquad \bar{\epsilon}_a = - H^s \phi^b_0
 b_{ba}. 
\label{eq:true_ansatz}
\end{align}
The constant $s$ will be determined later. 
We will see that \eqref{eq:true_ansatz} is the correct ansatz in due course.
Then the explicit variation of the relevant components are given by 
\begin{align}
\delta {\Isg}_{ab}
&= {\mathcal L}_\epsilon {\Isg}_{ab}
= H^s \Big[ 2s \phi_{0(a} \partial_{b)} H
	+ \delta_{ab} \phi_0^c \partial_c H \Big],
\notag \\
\delta {\IsB}_{ab}
&= {\mathcal L}_\epsilon {\IsB}_{ab} + 2 \partial_{[a} \bar{\epsilon}_{b]}
= H^s \phi_0^c {\IsH}_{abc} 
= H^s \phi_0^c \varepsilon_{abcd} \partial^d H, 
\notag \\
\delta {\Isphi}
&= {\mathcal L}_\epsilon {\Isphi}
= H^s \phi_0^a \partial_a {\Isphi}
= {1 \over 2} H^{s-1} \phi_0^a \partial_a H 
\label{eq:variation-phi}
\end{align}
where ${\IsH}_{\mu\nu\rho} = 3 \partial_{[\mu} {\IsB}_{\nu\rho]}$ and
we have used the BPS condition \eqref{eq:BPS_condition} in the
evaluation of $\delta {\IsB}_{ab}$.

Finally, fields that characterize the world-volume effective theory of a brane
are defined by promoting the constant moduli to functions on the
world-volume:
\begin{align}
\phi_0^a \to \phi^a ({\Isx}),
\label{eq:NS_promote}
\end{align}
where $\Isx$ is the coordinate of the brane world-volume.
We call these fields the fluctuation zero-modes of the five-branes.
In the following, we evaluate the equations \eqref{eq:NSNS_variations}
for the variations \eqref{eq:variation-phi} with the fields 
\eqref{eq:NS_promote} and determine the kinematics that governs the
fluctuation zero-modes.

\subsection{Zero-modes equation $\delta \mathcal{R}_{MN} = 0$}
We first evaluate the variation of the generalized Ricci tensor.
It is convenient to rewrite the components of the generalized Ricci
tensor as 
\begin{align}
{\mathcal R}_{\mu\nu}
&= {1 \over 2} (\text{${\Isg}$ eq.})_{\mu\nu}
	- ({\IsB}{\Isg}^{-1})_{(\mu|}{}^\beta (\text{${\IsB}$ eq.})_{|\nu)\beta}
	- {1 \over 2} ({\IsB}{\Isg}^{-1})_\mu{}^\alpha ({\IsB}{\Isg}^{-1})_\nu{}^\beta
		(\text{${\Isg}$ eq.})_{\alpha\beta}, 
\notag \\
{\mathcal R}_\mu{}^\nu
&= - {1 \over 2} {\Isg}^{\nu\beta} (\text{${\IsB}$ eq.})_{\mu\beta}
	- {1 \over 2} ({\IsB}{\Isg}^{-1})_\mu{}^\alpha {\Isg}^{\nu\beta}
		(\text{${\Isg}$ eq.})_{\alpha\beta},
\notag \\
{\mathcal R}^{\mu\nu}
&= - {1 \over 2} {\Isg}^{\mu\alpha} {\Isg}^{\nu\beta}
		(\text{${\Isg}$ eq.})_{\alpha\beta},
\label{eq:decomposition_gen_Ricci_tensor}
\end{align}
where we have defined the following expressions:
\begin{align}
(\text{${\Isg}$ eq.})_{\mu\nu}
&= \Italicsf{R}_{\mu\nu} - {1 \over 4} {\IsH}_{\mu\rho\sigma} {\IsH}_\nu{}^{\rho\sigma}
	+ 2 \nabla_\mu \nabla_\nu {\Isphi},
\notag \\
(\text{${\IsB}$ eq.})_{\mu\nu}
&= {1 \over 2} \nabla^\alpha {\IsH}_{\alpha\mu\nu} 
	- {\IsH}_{\alpha\mu\nu} \nabla^\alpha {\Isphi},
\notag \\
({\IsB}{\Isg}^{-1})_\mu{}^\nu
&= {\IsB}_{\mu\rho} {\Isg}^{\rho\nu},
\end{align}
where ${\Italicsf{R}}_{\mu\nu}$ and $\nabla_\mu$ are the Ricci tensor and 
the covariant derivative defined by ${\Isg}_{\mu\nu}$.
The derivation of the expressions \eqref{eq:decomposition_gen_Ricci_tensor} is
found in Appendix \ref{sct:detailed_calculations}.
Then, the variations of the generalized Ricci tensor become
\begin{align}
\delta {\mathcal R}_{\mu\nu}
&= {1 \over 2} \delta (\text{${\Isg}$ eq.})_{\mu\nu}
	- ({\IsB}{\Isg}^{-1})_{(\mu|}{}^\rho \delta (\text{${\IsB}$ eq.})_{|\nu)\rho}
	- {1 \over 2} ({\IsB}{\Isg}^{-1})_\mu{}^\rho ({\IsB}{\Isg}^{-1})_\nu{}^\sigma
		\delta (\text{${\Isg}$ eq.})_{\rho\sigma}, 
\notag \\
\delta {\mathcal R}_\mu{}^\nu
&= - {1 \over 2} {\Isg}^{\nu\rho} \delta (\text{${\IsB}$ eq.})_{\mu\rho}
- {1 \over 2} {\Isg}^{\nu\rho} ({\IsB}{\Isg}^{-1})_\mu{}^\sigma \delta 
	(\text{${\Isg}$ eq.})_{\rho\sigma},
\notag \\
\delta {\mathcal R}^{\mu\nu}
&= - {1 \over 2} {\Isg}^{\mu\rho} {\Isg}^{\nu\sigma} \delta (\text{${\Isg}$ eq.})_{\rho\sigma}.
\end{align}
Therefore, we find that the equations for the zero-modes are given by 
\begin{align}
\delta (\text{${\Isg}$ eq.})_{\mu\nu} = 0, \qquad
\delta (\text{${\IsB}$ eq.})_{\mu\nu} = 0.
\end{align}

\paragraph{Variation $\delta (\text{\rm ${\Isg}$ eq.})$}
We first analyze the variation  $\delta (\text{${\Isg}$ eq.})_{\mu\nu}$.
This is explicitly given by 
\begin{align}
\delta (\text{${\Isg}$ eq.})_{\mu\nu}
&= \delta \Italicsf{R}_{\mu\nu}
	- {1 \over 4} \delta ({\IsH}_{\mu\rho\sigma} {\IsH}_\nu{}^{\rho\sigma})
	+ 2 \delta (\nabla_\mu \nabla_\nu {\Isphi}).
\label{eq:variation-geq}
\end{align}
In the following, we write down the equation \eqref{eq:variation-geq} by
decomposing the indices into the world-volume and the transverse directions.
Using the variations \eqref{eq:variation-phi}, the non-zero variations of
the Christoffel symbol defined by ${\Isg}_{\mu\nu}$ for the localized DFT monopole 
\eqref{eq:DFT_monopole_ansatz} are found to be 
\begin{align}
\delta {\IsGamma}^m_{ab}
&= - {1 \over 2} H^s \Big[
	2 s (\partial^m \phi_{(a}) \partial_{b)} H
	+ \delta_{ab} (\partial^m \phi^d) \partial_d H \Big], 
\notag \\
\delta {\IsGamma}^c_{m b}
&= {1 \over 2} H^{s-1} \Big[ 
s \Big(
	(\partial_m \phi^c) \partial_b H
	+ (\partial_m \phi_b) \partial^c H \Big)
+ \delta^c_b (\partial_m \phi^d) \partial_d H \Big], 
\notag \\
\delta {\IsGamma}^c_{ab}
&= {1 \over 2} \Big[ 
2(s-1) H^{s-2} \Big( 
	s \phi^c \partial_a H \partial_b H
	+ \delta^c_{(a|} \phi^d \partial_d H \partial_{|b)} H \Big)
+ s H^{s-2} \Big( 
	\delta_{ab} \phi^c (\partial H)^2 
	- 2 \phi_{(a} \partial_{b)} H \partial^c H \Big) 
\notag \\
& \hspace{8mm}
+ H^{s-2} \delta_{ab} \phi^d \partial_d H \partial^c H 
+ H^{s-1} \Big( 
	2s \phi^c \partial_a \partial_b H
	+ 2 \delta^c_{(a|} \phi^d \partial_d \partial_{|b)} H 
	- \delta_{ab} \phi^d \partial_d \partial^c H
	\Big) 
\Big],
\label{eq:variation-Christoffel-abc}
\end{align}
where $m,n, \ldots = 0,1, \ldots, 5$ and $a,b, \ldots = 6, \ldots, 9$
are the world-volume and the transverse directions, respectively.
Then, we find the variations of the Ricci tensor are given by 
\begin{align}
\delta \Italicsf{R}_{mn}
&= - (s+2) H^{s-1} (\partial_m \partial_n \phi^c) \partial_c H,
\notag \\
\delta \Italicsf{R}_{mb}
&= - {1 \over 2} H^{s-2} \Big[ 
	(s^2 + s - 3) (\partial_m \phi^c) \partial_c H \partial_b H
	- s(s+1) (\partial_m \phi_b) (\partial H)^2 \Big]
\notag \\
& \quad
- {1 \over 2} H^{s-1} \Big[ 
	(s+3) (\partial_m \phi^c) \partial_c \partial_b H
	- s (\partial_m \phi_b) \square H \Big],
\notag \\
\delta \Italicsf{R}_{ab}
&= - {1 \over 2} H^s \Big[
	2s (\partial_m \partial^m \phi_{(a}) \partial_{b)} H
	+ \delta_{ab} (\partial_m \partial^m \phi^c) \partial_c H \Big]
\notag \\
& \quad 
+ {1 \over 2} \Big[ 
	H^{s-2} \delta_{ab} \phi^c \partial_c H \square H
	- 2s H^{s-2} \phi_{(a} \partial_{b)} H \square H
	- H^{s-1} \delta_{ab} \phi^c \partial_c \square H \Big]
\notag \\
& \quad
+ 3(s-1) H^{s-3} \phi^c \partial_c H \partial_a H \partial_b H
- H^{s-1} \phi^c \partial_c \partial_a \partial_b H
\notag \\
& \quad
+ H^{s-2} \Big[ 
	\phi^c \partial_c H \partial_a \partial_b H
	- (2s-3) \phi^c \partial_c \partial_{(a} H \partial_{b)} H \Big]
\label{eq:delta-R-ab}
\end{align}
Here we have defined $(\partial H)^2 = \delta^{ab} \partial_a H
\partial_b H$.
We next evaluate the variation 
\begin{align}
\delta ({\IsH}_{\it 3}^{\,2})_{\mu\nu}
=\; &
\delta ({\IsH}_{\mu\rho\sigma} {\IsH}_\nu{}^{\rho\sigma})
\notag \\
=\; & {\Isg}^{\rho\tau} {\Isg}^{\sigma\kappa} (\delta {\IsH}_{\mu\rho\sigma}) {\IsH}_{\nu\tau\kappa}
+ {\Isg}^{\rho\tau} {\Isg}^{\sigma\kappa} {\IsH}_{\mu\rho\sigma} (\delta {\IsH}_{\nu\tau\kappa}) 
+ 2 (\delta {\Isg}^{\rho\tau}) {\Isg}^{\sigma\kappa} {\IsH}_{\mu\rho\sigma} {\IsH}_{\nu\tau\kappa}.
\end{align}
Since we have only the following non-zero components for the localized DFT monopole,
\begin{align}
\delta {\IsH}_{mab}
&= H^s (\partial_m \phi^c) \varepsilon_{abcd} \partial^d H, 
\notag \\
\delta {\IsH}_{abc}
&= s H^{s-1} \phi^d \Big( 
	\varepsilon_{abde} \partial_c H
	+ \varepsilon_{bcde} \partial_a H 
	+ \varepsilon_{cade} \partial_b H
	\Big) \partial^e H
\notag \\
& \quad
+ H^s \phi^d \Big(
	\varepsilon_{abde} \partial^e \partial_c H
	+ \varepsilon_{bcde} \partial^e \partial_a H
	+ \varepsilon_{cade} \partial^e \partial_b H \Big),
\end{align}
the non-zero components of the variations are found to be 
\begin{align}
\delta ({\IsH}_{\it 3}^{\,2})_{ma}
&= 2 H^{s-2} \Big( 
	(\partial_m \phi_a) (\partial H)^2
	- (\partial_m \phi^c) \partial_c H \partial_a H \Big), 
\notag \\
\delta ({\IsH}_{\it 3}^{\,2})_{ab}
&= 4 H^{s-3} \Big( 
	s\phi_{(a} \partial_{b)} H (\partial H)^2 
	- \delta_{ab} \phi^c \partial_c H (\partial H)^2 
	- (s-1) \phi^c \partial_c H \partial_a H \partial_b H \Big)
\notag \\
& \quad
+ 4H^{s-2} \Big( 
	\delta_{ab} \phi^c \partial_c \partial_d H \partial^d H
	- \phi^c \partial_c \partial_{(a} H \partial_{b)} H
	+ \phi_{(a} \partial_{b)} H \square H
	- \delta_{ab} \phi^c \partial_c H \square H \Big).
\end{align}
Finally, we evaluate the variation 
\begin{align}
\delta (\nabla_\mu \nabla_\nu {\Isphi})
&= \partial_\mu \partial_\nu (\delta {\Isphi})
	- (\delta {\IsGamma}^\rho_{\mu\nu}) \partial_\rho {\Isphi} 
	- {\IsGamma}^\rho_{\mu\nu} \partial_\rho (\delta {\Isphi}).
\end{align}
By using \eqref{eq:variation-phi} and
\eqref{eq:variation-Christoffel-abc}, we find the non-zero components of
the above variation are 
\begin{align}
\delta (\nabla_m \nabla_n {\Isphi})
&= {1 \over 2} H^{s-1} (\partial_m \partial_n \phi^c) \partial_c H,
\notag \\
\delta (\nabla_m \nabla_a {\Isphi})
&= {1 \over 4} \Big[ H^{s-2} \Big(
	(s-3) (\partial_m \phi^c) \partial_c H \partial_a H
	- s (\partial_m \phi_a) (\partial H)^2 \Big)
	+ 2 H^{s-1} (\partial_m \phi^c) \partial_c \partial_a H \Big],
\notag \\
\delta (\nabla_a \nabla_b {\Isphi})
&= {1 \over 2} \Big( 
H^{s-2} \delta_{ab} \phi^c \partial_c \partial_d H \partial^d H
+ 2(s-2) H^{s-2} \phi^c \partial_c \partial_{(a} H \partial_{b)} H
- H^{s-2} \phi^c \partial_c H \partial_a \partial_b H
\notag \\
& \hspace{12mm}
- 4(s-1) H^{s-3} \phi^c \partial_c H \partial_a H \partial_b H 
- H^{s-3} \delta_{ab} \phi^c \partial_c H (\partial H)^2 
\notag \\
& \hspace{12mm}
+ s H^{s-3} \phi_{(a} \partial_{b)} H (\partial H)^2  
+ H^{s-1} \phi^c \partial_c \partial_a \partial_b H
\Big).
\end{align}
Collecting all together, we find that the equation of the zero-modes $\delta (\text{${\Isg}$
eq.})_{\mu \nu} = 0$ results in the following form: 
\begin{align}
0 = \delta (\text{${\Isg}$ eq.})_{mn}
&= - (s+1) H^{s-1} (\partial_m \partial_n \phi^c) \partial_c H,
\notag \\
0= \delta (\text{${\Isg}$ eq.})_{ma}
&= {1 \over 2} (s+1)(s-1) H^{s-2} \Big( 
	(\partial_m \phi_a) (\partial H)^2
	- (\partial_m \phi^c) \partial_c H \partial_a H \Big)
\notag \\
& \quad
- {1 \over 2} H^{s-1} \Big[ 
	(s+1) (\partial_m \phi^c) \partial_c \partial_a H
	- s (\partial_m \phi_a) \square H \Big],
\notag \\
0 = \delta (\text{${\Isg}$ eq.})_{ab}
&= - {1 \over 2} H^s \Big[
	2s (\partial_m \partial^m \phi_{(a}) \partial_{b)} H
	+ \delta_{ab} (\partial_m \partial^m \phi^c) \partial_c H \Big]
\notag \\
& \quad 
+ {1 \over 2} \Big[ 
	3 H^{s-2} \delta_{ab} \phi^c \partial_c H \square H
	- 2(s+1) H^{s-2} \phi_{(a} \partial_{b)} H \square H
	- H^{s-1} \delta_{ab} \phi^c \partial_c \square H \Big].
\end{align}
Since $H$ is a harmonic function $\square H = 0$, we find $s=-1$ and
the scalar fields $\phi^a$ satisfy the Klein-Gordon equation 
$\partial_m \partial^m \phi^a = 0$ in the six-dimensional world-volume.

\paragraph{Variation $\delta (\text{\rm ${\IsB}$ eq.})$}
We next evaluate the variation 
\begin{align}
\delta (\text{${\IsB}$ eq.})_{\mu\nu}
&= {1 \over 2} \delta (\nabla^\alpha {\IsH}_{\alpha\mu\nu})
	- \delta ({\IsH}_{\alpha\mu\nu} \nabla^\alpha {\Isphi}).
\end{align}
One easily finds that $\delta (\text{${\IsB}$ eq.})_{mn} 
= \delta (\text{${\IsB}$ eq.})_{ma} = 0$ for the localized DFT monopole.
Therefore only non-zero component in this variation is 
\begin{align}
\delta (\text{${\IsB}$ eq.})_{ab}
&= {1 \over 2} H^s \varepsilon_{abcd} (\partial_m \partial^m \phi^c) \partial^d H
+ {1 \over 2} H^{s-1} \phi^c \varepsilon_{abcd} \partial^d \square H
\notag \\
& \quad
+ (s-2) H^{s-2} \phi^c \Big( 
	\varepsilon_{ab[c|e} \partial^e \partial_{|d]} H 
	+ \varepsilon_{[a|cde} \partial^e \partial_{|b]} H \Big) \partial^d H.
\end{align}
By using the condition $\square H = 0$ and the relation
\begin{align}
\varepsilon_{[abc|e} \partial^e \partial_{|d]} H
&= {1 \over 2} \Big( 
	\varepsilon_{[a|cde} \partial^e \partial_{|b]} H 
	+ \varepsilon_{ab[c|e} \partial^e \partial_{|d]} H \Big) 
= 0,
\end{align}
which originates from the Bianchi identity $\partial_{[a} {\IsH}_{bcd]} = 0$, 
we have the zero-modes equation,
\begin{align}
0 = 
\delta (\text{${\IsB}$ eq.})_{ab}
&= {1 \over 2} H^s \varepsilon_{abcd} (\partial_m \partial^m \phi^c) \partial^d H.
\end{align}
This is again satisfied by the Klein-Gordon scalar fields $\phi^a$.

\subsection{Zero-modes equation $\delta \mathcal{R} = 0$}
We next evaluate the variation $\delta \mathcal{R} = 0$.
The generalized Ricci scalar in the parametrization
\eqref{eq:monopole(3.3)}
is given by 
\begin{align}
{\mathcal R}
=& 
{1 \over 4} {\Isg}^{\mu\nu} \partial_\mu {\Isg}^{\rho\sigma} \partial_\nu {\Isg}_{\rho\sigma}
- {1 \over 2} {\Isg}^{\mu\nu} \partial_\nu {\Isg}^{\rho\sigma} \partial_\sigma {\Isg}_{\mu\rho}
- {1 \over 12} {\IsH}_{\mu\nu\rho} {\IsH}^{\mu\nu\rho}
\notag \\
& \quad
+ 4 {\Isg}^{\mu\nu} \partial_\mu \partial_\nu {\Isd} 
- \partial_\mu \partial_\nu {\Isg}^{\mu\nu} 
- 4 {\Isg}^{\mu\nu} \partial_\mu {\Isd} \, \partial_\nu {\Isd}
+ 4 \partial_\mu {\Isg}^{\mu\nu} \partial_\nu {\Isd}.
\label{eq:DFTRicciScalar-v2}
\end{align}
It is convenient to rewrite the generalized Ricci scalar in terms of the
well-known geometric quantities.
A tedious rearrangement leads to the following expression
\begin{align}
{\mathcal R}
&= 
\Italicsf{R} + 4 (\nabla^\mu \nabla_\mu {\Isphi} - (\nabla {\Isphi})^2)
- {1 \over 12} {\IsH}_{\mu\nu\rho} {\IsH}^{\mu\nu\rho},
\label{eq:gen_Ricci_scalar_sc}
\end{align}
where we have used the relation $\partial_\mu {\Isd} = \partial_\mu
{\Isphi} - {1 \over 2} {\IsGamma}^\rho_{\mu\rho}$ and $\Italicsf{R}$ is the
ordinary Ricci scalar defined by 
\begin{align}
\Italicsf{R}
&= {\Isg}^{\mu\nu} \Italicsf{R}_{\mu\nu}
= {\Isg}^{\mu\nu} \Big( \partial_\rho {\IsGamma}^\rho_{\mu\nu}
	- {\IsGamma}^\sigma_{\rho\mu} {\IsGamma}^\rho_{\nu\sigma}
	- \partial_\mu {\IsGamma}^\rho_{\nu\rho}
	+ {\IsGamma}^\sigma_{\rho\sigma} {\IsGamma}^\rho_{\mu\nu} \Big).
\end{align}
Having an insight that \eqref{eq:gen_Ricci_scalar_sc} is the same form with the equation
of motion for the dilaton $\Phi$ in the NSNS sector in supergravity, 
it is further convenient to rewrite the generalized Ricci scalar as 
\begin{align}
\mathcal{R}
&= {\Isg}^{\mu\nu} (\text{${\Isg}$ eq.})_{\mu\nu}
+ 2 \nabla^\mu \nabla_\mu {\Isphi} - 4 (\nabla {\Isphi})^2
+ {1 \over 6} {\IsH}_{\mu\nu\rho} {\IsH}^{\mu\nu\rho}.
\end{align}
Then, the variation of the generalized Ricci scalar is given by 
\begin{align}
\delta \mathcal{R}
&= (\delta {\Isg}^{\mu\nu}) (\text{${\Isg}$ eq.})_{\mu\nu}
+ {\Isg}^{\mu\nu} \delta (\text{${\Isg}$ eq.})_{\mu\nu}
\notag \\
& \quad
+ 2 \delta ({\Isg}^{\mu\nu} \nabla_\mu \nabla_\nu {\Isphi}) 
- 4 \delta ({\Isg}^{\mu\nu} \nabla_\mu {\Isphi} \nabla_\nu {\Isphi})
+ {1 \over 6} \delta ({\Isg}^{\mu\nu} {\Isg}^{\rho\sigma} {\Isg}^{\tau\kappa} 
	{\IsH}_{\mu\rho\tau} {\IsH}_{\nu\sigma\kappa}).
\end{align}
With the help of the results in the previous paragraph, each term is evaluated as  
\begin{align}
& (\delta {\Isg}^{\mu\nu}) (\text{${\Isg}$ eq.})_{\mu\nu}
= 0,
\notag \\
& {\Isg}^{\mu\nu} \delta (\text{${\Isg}$ eq.})_{\mu\nu}
= - (2s + 3) H^{s-1} (\partial_m \partial^m \phi^c) \partial_c H,
\notag \\
& 2 \delta ({\Isg}^{\mu\nu} \nabla_\mu \nabla_\nu {\Isphi}) 
= H^{s-1} (\partial_m \partial^m \phi^c) \partial_c H,
\notag \\
& - 4 \delta ({\Isg}^{\mu\nu} \nabla_\mu {\Isphi} \nabla_\nu {\Isphi})
= 3 H^{s-4} \phi^c \partial_c H (\partial H)^2
	- 2 H^{s-3} \phi^c \partial_c \partial_a H \partial^a H,
\notag \\
& {1 \over 6} \delta ({\Isg}^{\mu\nu} {\Isg}^{\rho\sigma} {\Isg}^{\tau\kappa} 
	{\IsH}_{\mu\rho\tau} {\IsH}_{\nu\sigma\kappa})
= - 3 H^{s-4} \phi^c \partial_c H (\partial H)^2 
	+ 2 H^{s-3} \phi^c \partial_c \partial_a H \partial^a H.
\end{align}
Collecting all together, we obtain the equation for the translational zero-modes,
\begin{align}
0 = 
\delta \mathcal{R}
&= - 2 (s+1) H^{s-1} (\partial_m \partial^m \phi^c) \partial_c H.
\end{align}
From this expression, we again find that $\phi^a$ are scalar fields satisfying the
Klein-Gordon equation in the world-volume and $s = -1$. This is consistent with the
results obtained in the analysis of $\delta \mathcal{R}_{MN} = 0$.

In summary, we have the translational zero-modes given by 
\begin{align}
\epsilon^a = H^{-1} \phi^a (x), \qquad \bar{\epsilon}_a = - H^{-1}
 \phi^b (x) b_{ba},
\end{align}
where the scalar fields $\phi^a$ satisfies the Klein-Gordon equation in the
six-dimensional brane world-volume.
These zero-modes are Nambu-Goldstone modes associated with the
spontaneous breaking of the translational symmetry along the transverse
doubled space to five-branes.

\section{Tensor zero-modes of five-branes} \label{sct:tensor}
In the previous section, we determined the zero-modes associated with
the spontaneous breaking of the translational symmetry in the doubled
space. 
In this section, we determine the tensor zero-modes associated with the 
spontaneous breaking of the gauge symmetry in the RR sector.

\subsection{Zero-modes in the RR sector}
The localized DFT monopole \eqref{eq:DFT_monopole_ansatz} breaks the
gauge symmetry of the RR potential in \eqref{eq:DFT_gauge_transformations}:
\begin{align}
\delta_{\lambda} \chi = \slashed{\del} \lambda,
\label{eq:RR_gauge_transformation}
\end{align}
where $\lambda$ is an $O(D,D)$ spinor parameter.
We examine a fluctuation $\delta_\lambda \chi$ which is given by the gauge
transformation \eqref{eq:RR_gauge_transformation} in the localized DFT monopole
solution. 
In order that, 
we employ the ansatz of the Nambu-Goldstone modes
\begin{align}
\lambda = e^{\frac{1}{2} b_{ab} \psi^{a} \psi^{b}} H^t ({\Isy})
 \bar{\lambda},
\label{eq:RR_ansatz}
\end{align}
where the $p$-form parameters in the constant $O(D,D)$ spinor $\bar{\lambda}$ extend along the
world-volume directions.
The overall exponential factor guarantees that 
$H^t ({\Isy}) \bar{\lambda}$
is the genuine zero-mode which is invariant under the NSNS $B$-field
gauge transformation \cite{Hohm:2011dv}.
Here $H({\Isy})$ is the harmonic function defined in the localized DFT
monopole solution and $t$
is a real number which will be determined later.

Since the localized DFT monopole solution \eqref{eq:DFT_monopole_ansatz} is given within the
solution to the strong constraint, we assume $\bar{\del}^{\mu} = 0$ in the following.
Then the fluctuation \eqref{eq:RR_ansatz} becomes 
\begin{align}
\delta_{\lambda} \chi 
=& \ 
e^{\frac{1}{2} b \psi \psi}
\left[
\frac{1}{2} H^t \del_c b_{ab} \psi^{ab} + t H^{t-1} \del_c H 
\right] \psi^c \bar{\lambda}.
\label{eq:RR_gauge_transformation_zero-mode}
\end{align}
Here we have used the fact that $b$ has only the transverse components
in the solution. We have also introduced the notation $b \psi \psi = b_{ab} \psi^a \psi^b$.
In the following, we also use the notation such as $\psi^{\mu \nu \rho \cdots} = \psi^{\mu} \psi^{\nu} \psi^{\rho} \cdots$.

We next promote the constant parameter to the field in the world-volume:
\begin{align}
\bar{\lambda} \to \lambda ({\Isx}).
\label{eq:RR_promote}
\end{align}
The component expansion of the $O(D,D)$ spinor is given by 
\begin{align}
\lambda ({\Isx}) = \sum_{p=1}^{D} \frac{1}{(p-1)!} \lambda_{m_1 \cdots m_{p-1}} ({\Isx}) 
\psi^{m_1 \cdots m_{p-1}} | 0 \rangle.
\end{align}
Now we substitute the fluctuation into the equation of motion.
The only relevant equation of motion for the RR zero-modes is 
\begin{align}
\slashed{\del}
\left(
\mathcal{K} \slashed{\del} \delta_{\lambda} \chi
\right) = 0.
\label{eq:RReom}
\end{align}
Here $\mathcal{K} = C^{-1} \mathbb{S}_{\IsH}$.
Substituting the fluctuation into the equation of motion
and performing tedious calculations, we find
(detailed calculations are found in Appendix~\ref{sct:detailed_calculations}),
\begin{align}
&
\Bigg[
(t^2-1) H^{t-1} \del_a H \del_a H
\Bigg] 
\sum_{p} F^{l n_1 \cdots n_{p-1}}  
\psi^{l n_1 \cdots n_{p-1}} | 0 \rangle 
\notag \\
& 
-
\Bigg[
\frac{1}{2} H^{t-1} \del_c b_{ef} \psi^{efc} + t H^t \del_c H \psi^c
\Bigg] 
\sum_p p \, \del_m F^{m n_1 \cdots n_{p-1}} 
 \psi^{n_1 \cdots n_{p-1}} |0 \rangle
= 0,
\label{eq:RR_zero-modes1}
\end{align}
where we have defined
\begin{align}
F^{m_1 m_2 \cdots m_p} =
p! \del^{[m_1} \lambda^{m_2 \cdots m_p]}.
\end{align}
Therefore, we conclude that the zero-mode equation is satisfied 
when $t = \pm 1$ and 
\begin{align}
\del_m F^{m n_1 \cdots n_{p-1}} = 0.
\label{eq:maxwell_eq}
\end{align}
The condition \eqref{eq:maxwell_eq} is nothing but the Maxwell equation
of the $p$-forms $\lambda$ in the six-dimensional world-volume.

\subsection{Normalizability}
In order to determine the value of $t$, we next examine the normalizability of the RR zero-modes.
To this end, we evaluate the effective action of the RR zero-modes.
This is obtained in the RR sector of the DFT action \eqref{eq:DFT_action}.
We introduce the fluctuation $\delta_\lambda \chi$ given by the gauge
transformation \eqref{eq:RR_gauge_transformation_zero-mode}, with the
promoted gauge parameter \eqref{eq:RR_promote} in the action.
Then the relevant term in the action becomes
\begin{align}
S_{\mathrm{DFT,RR}} = \frac{1}{4} \int \! {\dop}^D \tilde{x} \int \! {\dop}^D x \, 
(\slashed{\del} \delta_{\lambda} \chi)^{\dagger} \mathbb{S}_{\IsH} \,
 \slashed{\del} \delta_{\lambda} \chi,
\end{align}
where $\mathbb{S}_{\IsH}$  
is the $O(D,D)$ spinor representation
of the generalized metric for the localized DFT monopole solution \eqref{eq:DFT_monopole_ansatz}. 
Once again, calculations give the following result 
(details are found in Appendix~\ref{sct:detailed_calculations}),
\begin{align}
(\slashed{\del} \delta_{\lambda} \chi)^{\dagger} 
\mathbb{S}_{\IsH} \,
 (\slashed{\del} \delta_{\lambda} \chi) 
=& \ 
- \frac{1}{4} 
H^{2t-1} 
\del_c b_{ef} \del_{c'} b_{e'f'} \sum_{p}
 f_{n m_1 \cdots m_{p-1}} \sum_{q} f_{n'
 m'_1 \cdots m'_{q-1}}
\eta^{n'l'} \eta^{m'_1 n'_1} \cdots
 \eta^{m'_{q-1} n'_{q-1}} 
\notag \\
& \ 
\qquad \qquad 
\times 
\langle 0 | \psi_{m_{p-1} \cdots m_1} \psi_{cfe} 
\psi^{e'f'c'} \psi^{l' n'_1 \cdots n'_{q-1}} |0
 \rangle
\notag \\
& \ 
- t^2 H^{2t-1} \del_c H \del_{c'} H 
\sum_{p}
 f_{n m_1 \cdots m_{p-1}} \sum_{q} f_{n'
 m'_1 \cdots m'_{q-1}}
\eta^{n'l'} \eta^{m'_1 n'_1} \cdots
 \eta^{m'_{q-1} n'_{q-1}} 
\notag \\
& \ 
\qquad \qquad 
\times
\langle 0 | \psi_{m_{p-1} \cdots m_1} \psi_{c} 
\psi^{c'} \psi^{l' n'_1 \cdots n'_{q-1}} |0
 \rangle.
\label{eq:effective_RR}
\end{align}
The doubled volume factor is decomposed into the world-volume and the
transverse directions $\int \! {\dop}^D \tilde{x} {\dop}^D x = \int_{\mathrm{wv.}} \! {\dop}^6
\tilde{x} {\dop}^6 x \int_{\mathrm{trans.}} \!\!\! {\dop}^4 \tilde{x} {\dop}^4x$.
Here ``wv.''\ and ``trans.''\ mean the world-volume and the transverse
spaces, respectively.
The volume factors $\int_{\mathrm{wv.}} \! {\dop}^6 \tilde{x}$ and
$\int_{\mathrm{trans.}} \! {\dop}^4 \tilde{x}$ give finite values for the compactified torus.
In order to obtain the world-volume theory, we now integrate out the
transverse directions that survives the strong constraint:
\begin{align}
\int_{\text{trans.}} \!\! {\dop}^4 x \, 
(\slashed{\del} \delta_{\lambda} \chi)^{\dagger} 
\mathbb{S}_{\IsH}\,
 (\slashed{\del} \delta_{\lambda} \chi) 
\end{align}
We evaluate the integrals of the first and the second terms in \eqref{eq:effective_RR}.
The first term gives
\begin{align}
&
\int_0^{\infty} \! r^3 {\dop}r \, 
H^{2t-1} \del_{[a} b_{bc]} \del_{[d} b_{ef]} 
\langle 0| \psi_{abc} \psi^{def} |0 \rangle
\notag \\
\sim& \ 
\int_0^{\infty} \! r^3 {\dop}r \, 
H^{2t-1} \del_g H \del_h H \varepsilon_{abcg} \varepsilon_{defh} \,
\delta_{a} {}^{[d} \delta_b {}^e \delta_c {}^{f]}
\notag \\
\sim& \ 
\int_0^{\infty} \! r^3 {\dop}r \, 
H^{2t-1} \del_a H \del_a H.
\end{align}
Here we have omitted the irrelevant overall factors.
For the localized DFT monopole of the codimension four, we find
$H = c_1 + c_2 r^{-2}$ 
and the relevant integral is given by 
\begin{align}
\int_0^{\infty} \! {\dop}r \, r^{-3} 
\left(
c_1 + c_2 r^{-2}
\right)^{2t-1}.
\label{eq:RR_normalizable_integral}
\end{align}
This is finite only when $t = -1$.
For $t = 1$, the integral diverges at $r = 0$.
Similarly, the second term in \eqref{eq:effective_RR} also gives the integral
\eqref{eq:RR_normalizable_integral}, 
giving a finite result only when $t = -1$.

For the DFT monopole of the codimension three, we have 
$H = c_1 + c_2 r^{-1}$
and the relevant integral becomes
\begin{align}
\int_0^{\infty} \! {\dop}r \, 
r^{-2} 
\left(
c_1 + c_2 r^{-1}
\right)^{2t-1}.
\end{align}
This again gives a finite result only when $t = -1$.
For the five-brane of codimension two, we have
\begin{align}
H = c_1 - c_2 \ln r,
\end{align}
where we have assumed $c_1, c_2 > 0$.
The integral becomes 
\begin{align}
I_t = \int_0^{\infty} \! {\dop}r \,
r^{-1} 
\left(
c_1 - c_2 \ln r
\right)^{2t-1}.
\end{align}
This diverges both for $t = \pm 1$. However, this is an artificial
phenomenon originating from the fact that the codimension two brane is
ill-defined as a stand alone object.
Indeed, if we introduce a cutoff at $r = M$, this gives a finite result only when $t = -1$:
\begin{align}
I_{t=-1} = \frac{1}{2 c_2 ( c_1 - c_2 \ln M)^2}.
\end{align}
For the DFT monopole of the codimension one, we have,
\begin{align}
H = c_1 + c_2 |x|,
\end{align}
where we have assumed that $c_1, c_2 > 0$.
The relevant integral becomes 
\begin{align}
\int_{-\infty}^{\infty} \! {\dop}x \, 
(c_1 + c_2 |x|)^{2t-1}.
\end{align}
This is finite only when $t = -1$.
Therefore, we conclude that $t = -1$ is the correct value of the normalizable zero-modes.

The same analysis is also applied in the NSNS sector.
By using the explicit variations of the fields
\begin{align}
\delta {\Isg}_{ab} 
= H^{-1} \Big( \delta_{ab} \phi^c \partial_c H - 2 \phi_{(a}
 \partial_{b)} H \Big), 
\quad 
\delta {\IsB}_{ab}
= H^{-1} \phi^c \varepsilon_{abcd} \partial^d H, 
\quad 
\delta {\Isphi}
= {1 \over 2} H^{-2} \phi^a \partial_a H,
\end{align}
we find that the variation of the DFT action results in
\begin{align}
\delta^2 S_{\text{DFT,NSNS}} =  2 \int \! {\dop}^{2D} X \, H^{-3}
\del_a H \del_b H
(\phi^a 
\Box \phi^b).
\end{align}
Then the relevant integral results in
\begin{align}
\int_{\text{trans.}} \!\! {\dop}^4 x \, H^{-3} \del_a H \del_b H = \frac{\pi^2
 c_2}{2 c_1^2} \delta_{ab}, 
\end{align}
giving a finite value for the harmonic function $H = c_1 + c_2 r^{-2}$.
This implies that the scalar part of the effective action of the five-brane
is given by the desired form:
\begin{align}
S \sim \int_{\text{wv.}} \! {\dop}^6 x \, \phi^a \Box \phi^a.
\end{align}
For the five-branes of lower codimensions, the calculations are the
same with the ones in the RR sector.

\subsection{Self-duality constraint}
Finally, we analyze the self-duality constraint of the $O(D,D)$ spinor $\chi$.
Since we use the democratic formulation of the RR sector, we need to 
impose the self-duality constraint in order to rewrite the higher rank forms by the lower ones.
The self-duality constraint is expressed as 
\begin{align}
\slashed{\del} \chi = - \mathcal{K} \slashed{\del} \chi.
\label{eq:sd1}
\end{align}
We derive the corresponding constraints for the fluctuations.
By substituting the shift 
$\chi \to \chi_0 + \delta \chi = 0 + \delta_{\lambda} \chi$ 
into the self-duality constraint, we have the following relation
(the detailed calculations are found in Appendix~\ref{sct:detailed_calculations})
\begin{align}
&
\sum_{p=0}^6 \frac{1}{(6-p)!} \varepsilon_{n m_1 \cdots m_{p-1} k_1 \cdots k_{6-p}}
\eta_{k_1 l_1} \cdots \eta_{k_{6-p} l_{6-p}} F_{n m_1 \cdots m_{p-1}}
\psi^{l_1 \cdots l_{6-p}}
=
- t \sum_{p=0}^6 F_{l_1 \cdots l_p} \psi^{l_1 \cdots l_{p}},
\notag 
\\
&
\sum_{p=0}^6 \frac{1}{(6-p)!} \varepsilon_{n m_1 \cdots m_{p-1} k_1 \cdots k_{6-p}}
\eta_{k_1 l_1} \cdots \eta_{k_{6-p} l_{6-p}} F_{n m_1 \cdots m_{p-1}}
\psi^{l_1 \cdots l_{6-p}}
=
- \frac{1}{t} \sum_{p=0}^6 F_{l_1 \cdots l_p} \psi^{l_1 \cdots l_{p}}.
\label{eq:sd_comp2}
\end{align}
We find that these conditions are consistent when $t = -1$ which precisely coincides with the result in the previous section.
For $t = -1$, these conditions are rewritten as 
\begin{align}
F_{m_1 \cdots m_{6-p}} = 
\frac{(-)^{\frac{(6-p)(6-p-1)}{2}}}{(6-p)!} \varepsilon_{m_1 \cdots m_{6-p}}
{}^{n_{1} \cdots n_{p}} F_{n_{1} \cdots n_p}.
\label{eq:6dsd}
\end{align}
Here $\varepsilon$ is the Levi-Civita symbol in six dimensions.
The indices are raised and lowered by the six-dimensional Lorentz metric $\eta^{m n}$.
We denote the condition \eqref{eq:6dsd} as 
\begin{align}
F^{(p-6)} = \pm *_6 F^{(p)}.
\label{eq:6dsd_hodge}
\end{align}
This is nothing but the self-duality relation for the zero-modes in six
dimensions.
Now we write down the condition \eqref{eq:6dsd_hodge} explicitly in cases
of type IIA and IIB supergravities. 

\paragraph{Type IIA case}
For the type IIA case, the relevant RR potentials are
\begin{align}
C^{(1)}, \ C^{(3)}, \ C^{(5)}, \ C^{(7)}, \ C^{(9)}.
\end{align}
The corresponding gauge transformations are given by
\begin{align}
\delta C^{(1)} = {\dop} \lambda^{(0)},
\quad  
\delta C^{(3)} = {\dop} \lambda^{(2)}, 
\quad  
\delta C^{(5)} = {\dop} \lambda^{(4)}, 
\quad  
\delta C^{(7)} = {\dop} \lambda^{(6)}, 
\quad  
\delta C^{(9)} = {\dop} \lambda^{(8)}.
\end{align}
Since the zero-modes associated with the RR gauge transformations
are defined in the six-dimensional world-volume, 
we have $\lambda^{(8)} = {\dop} \lambda^{(6)} = 0$.
Therefore, the non-zero field strengths for the zero-modes are given by 
\begin{align}
F^{(1)}, \quad F^{(3)}, \quad F^{(5)}.
\end{align}
Here $F^{(p+1)} = {\dop} \lambda^{(p)}$ and the exterior derivative is
defined in the six-dimensional world-volume.
The self-duality relations \eqref{eq:6dsd_hodge} are therefore 
\begin{align}
F^{(1)} = - *_6 F^{(5)},
\qquad 
F^{(3)} = - *_6 F^{(3)},
\qquad
F^{(5)} = - *_6 F^{(1)}.
\end{align}
The first and the last ones state that the degrees of freedom by 
$\lambda^{(4)}$ are given by $\lambda^{(0)}$.
The second one implies that the 2-form $\lambda^{(2)}$ is anti-self-dual (ASD).
Therefore, the net zero-modes are given by 
\begin{align}
\lambda^{(0)} \ (\text{scalar}), \quad \lambda^{(2)} \ (\text{ASD 2-form}).
\end{align}
These together with the fluctuation zero-modes $\phi^{a}, \, (a=6,7,8,9)$
precisely give the bosonic components of the six-dimensional $\mathcal{N} = (2,0)$ 
tensor multiplet.

\paragraph{Type IIB case}
For the type IIA case, the relevant RR potentials are given by 
\begin{align}
C^{(0)}, \ C^{(2)}, \ C^{(4)}, \ C^{(6)}, \ C^{(8)}, \ C^{(10)}.
\end{align}
The corresponding gauge transformations are 
\begin{align}
\delta C^{(0)} = 0,
\quad  
\delta C^{(2)} = {\dop} \lambda^{(1)}, 
\quad  
\delta C^{(4)} = {\dop} \lambda^{(3)}, 
\quad  
\delta C^{(6)} = {\dop} \lambda^{(5)}, 
\quad  
\delta C^{(8)} = {\dop} \lambda^{(7)},
\quad
\delta C^{(10)} = {\dop} \lambda^{(9)}.
\end{align}
The RR 0-form is gauge invariant.
Since $\lambda^{(p)}$ are defined in the six-dimensional world-volume, 
we have $\lambda^{(7)} = \lambda^{(9)} = 0$.
Therefore the field strengths of the zero-modes are given by 
\begin{align}
F^{(2)}, \quad F^{(4)}, \quad F^{(6)}.
\end{align}
The self-duality relations \eqref{eq:6dsd_hodge} read 
\begin{align}
F^{(2)} = *_6 F^{(4)},
\qquad 
F^{(4)} = *_6 F^{(2)},
\qquad 
F^{(6)} = 0.
\end{align}
The first and the second ones state that the degrees of freedoms of 
$\lambda^{(3)}$ are given by $\lambda^{(1)}$.
The last one implies $\lambda^{(5)}$ is non-dynamical in the five-brane world-volume.
Therefore the net zero-mode is 
\begin{align}
\lambda^{(1)} \ \text{(vector)}.
\end{align}
This together with the fluctuation zero-modes $\phi^{a}, \, (a=6,7,8,9)$
precisely give the bosonic components of the six-dimensional $\mathcal{N} = (1,1)$ vector multiplet.

\section{Effective theories of locally non-geometric five-branes}
\label{sct:winding}
In this section, we examine the world-volume effective theories of the five-branes in type II string theories.
In particular, we focus on the effective theories of locally non-geometric five-branes.
In the previous sections, we have determined the zero-modes of the five-branes.
There are the four fluctuation zero-modes $\phi^a \, (a=6,7,8,9)$ and the
tensor zero-modes in the RR sector.
There are two classes of the tensor zero-modes, namely, the 1-form $\lambda^{(1)}$ and a pair of the scalar and
the anti-self-dual 2-form $(\lambda^{(0)}, \lambda^{(2)})$ depending on
the possible RR potentials in type IIA and IIB supergravities.
They satisfy the desired field equations in the world-volumes and are organized into the 
bosonic sectors of the six-dimensional $\mathcal{N} = (1,1)$ vector and the $\mathcal{N} = (2,0)$ tensor multiplets.
These multiplets and the corresponding five-branes are summarized in Table~\ref{tb:wvSUSY_multiplets}.

\begin{table}[t]
\centering
\begin{tabular}{l||l|c|l|c}
6d supermultiplets & \multicolumn{4}{c}{Five-branes} 
\\
\hline
$\mathcal{N} = (2,0)$ tensor & IIA & NS5, Q5, SF5 & IIB & KKM, R5
\\
\hline
$\mathcal{N} = (1,1)$ vector & IIA & KKM, R5 & IIB & NS5, Q5, SF5
\end{tabular}
\caption{The five-branes in type II theories and the corresponding
 supermultiplets in the world-volumes. The KKM and SF5 stand for the
 KK-monopole and the space-filling five-brane.}
\label{tb:wvSUSY_multiplets}
\end{table}

We now examine the physical meaning of these zero-modes in each brane.
Before that, we note that there are two kinds of locally non-geometric objects in string theory.
One is the R-brane type that appear in Table \ref{tb:wvSUSY_multiplets},
the other is the localized KK-monopole type \cite{Berman:2014jsa}.
We first consider the R-brane type and then introduce the localized KK-monopole type.

\subsection{R-brane type}
The locally non-geometric objects of the R-brane type appear in the
T-duality orbit including the ordinary geometric branes.
In the following, we survey the effective theories of five-branes by 
starting from the familiar NS5-brane and applying the T-duality transformations.

\paragraph{NS5-brane and KK-monopole}
For the type IIA NS5-brane, the world-volume theory is characterized by the
six-dimensional $\mathcal{N} = (2,0)$ tensor multiplet.
The 0- and 2-form fields in the tensor multiplet correspond to the
zero-modes associated with the gauge transformations of the RR 1- and 3-forms.
Among other things, the 0-form field in the type IIA NS5-brane world-volume
is interpreted as the fluctuation along the M-circle.
For the type IIB NS5-brane, the world-volume theory is governed by the
six-dimensional $\mathcal{N} = (1,1)$ vector multiplet.
The 1-form in the vector multiplet corresponds to the zero-mode
associated with the gauge transformation of the RR 2-form.

For the zero-modes that come from the generalized Lie derivative, they
are decomposed into those by the diffeomorphism $\delta = \mathcal{L}_{\zeta}$ and the $B$-field
gauge transformation $\delta B = \dop \Lambda$ whose parameters are given by
\begin{align}
\zeta^a = \epsilon^a = H^{-1} \phi^a (x), \qquad \Lambda_a =
 \bar{\epsilon}_a = - H^{-1} \phi^b (x) b_{ba}.
\label{eq:NS5_zero-modes_parameters}
\end{align}
From this expression, the four scalar zero-modes $\phi^a$ in both of the NS5-branes 
are interpreted as
the geometric fluctuations of branes in the transverse directions.
Compared with the D-branes~\cite{Adawi:1998ta}, 
one notices that the modes $\phi^a$ necessarily present in the $B$-field gauge
transformation in the NS5-branes.

For the KK-monopoles, the transverse directions are given by the Taub-NUT
space which accommodates one isometry direction.
The T-duality transformation of the RR potentials increases or decreases their ranks.
We are interested in the RR potentials whose indices extend to the
world-volume directions of the five-branes.
The T-duality transformation of the RR $p$-form 
along the ${\Isy}^9$-direction results in the $(p+1)$-form whose increased
index is given by $a=9$. 
Correspondingly, the zero-modes of the NS5-branes associated with the gauge
transformations of the RR potentials become those
with the increased index given by the isometry direction $(a=9)$.
Therefore, the 0- and the 2-form zero-modes in the world-volume of the IIB KK-monopole
stem from the gauge transformations of the RR 2- and
4-forms whose gauge parameters contain the index $a=9$.
The same is true for the type IIA KK-monopole.

One of the four geometric zero-modes in the NS5-brane is T-dualized and
it turns out to be zero-modes associated with the gauge transformation of the NSNS
$B$-field. 
Explicitly, this is given by the parameters 
\begin{align}
\Lambda_a &= (\Lambda_i, \Lambda_9) = (\bar{\epsilon}_i, \epsilon^9)
= (H^{-1} \phi^9 A_i, H^{-1} \phi^9), 
\quad (i = 6,7,8).
\end{align}
Then we find that the corresponding gauge transformation becomes
\begin{align}
\delta B = {\dop} \Lambda = \phi^9 \eta,
\label{eq:B_gauge_TN}
\end{align}
where we have introduced the basis $\eta$ defined by
\begin{align}
\eta = H^{-1} {\dop}A + {\dop}H^{-1} \wedge ({\dop}\bar{\Isy}_9 + A).
\end{align}
One finds that this basis satisfies the self-duality condition in the
transverse four-dimensional Taub-NUT space
\begin{align}
\eta = *_4 \eta.
\end{align}
This is consistent with the known results \cite{Brill:1964zz, Pope:1978zx}.
These zero-modes are summarized in Table~\ref{tb:NS5KK5_RRmodes1}.
\begin{table}[t]
\centering
\begin{tabular}{l||c|c|c|c}
        & $\phi^6, \phi^7, \phi^8$ & $\phi^9$ & 0-form & 2-form
\\
\hline
IIA NS5 & geometric & geometric & $\delta C^{(1)} = {\dop} \lambda^{(0)} : \lambda^{(0)}$&
		 $\delta C^{(3)} = {\dop} \lambda^{(2)} : \lambda^{(2)}_{mn}$
\\
\hline
IIB KKM & geometric & self-dual $\delta B$ & $\delta C^{(2)} = {\dop}
	     \lambda^{(1)} : \lambda^{(1)}_{9}$ & $\delta C^{(4)}
		 = {\dop} \lambda^{(3)} : \lambda^{(3)}_{mn9}$
\\ \hline \hline
        & & & \multicolumn{2}{c}{1-form} 
\\
\hline
IIB NS5 & geometric & geometric 
	& \multicolumn{2}{c}{$\delta C^{(2)} = {\dop} \lambda^{(1)} : \lambda^{(1)}_m$} 
\\
\hline
IIA KKM & geometric & self-dual $\delta B$ 
	& \multicolumn{2}{c}{$\delta C^{(3)} = {\dop} \lambda^{(2)} : \lambda^{(2)}_{m9}$}
\end{tabular}
\caption{Physical interpretations of the zero-modes in the six-dimensional
 $\mathcal{N} = (2,0)$ tensor and the $\mathcal{N} = (1,1)$ vector
 multiplets. 
The cases of the NS5-brane and the KK-monopole.
The index structures of the gauge parameters are specified explicitly.}
\label{tb:NS5KK5_RRmodes1}
\end{table}

\paragraph{Q5-brane}
For the Q5-brane, there are two isometry directions to the transverse
space. Only the two fluctuation zero-modes correspond to the geometric modes in the transverse directions.
Two of the four fluctuation zero-modes in the NS5-brane are T-dualized 
giving rise to the two independent $B$-field gauge transformations.
These are explicitly given by
\begin{align}
\delta B = 
\phi^8 {\dop}H^{-1} \wedge {\dop}\bar{\Isy}_8 
+ \phi^9 {\dop}H^{-1} \wedge {\dop}\bar{\Isy}_9.
\label{eq:Q5_deltaB}
\end{align}
This is a generalization of the $B$-field gauge zero-mode
\eqref{eq:B_gauge_TN} in the Taub-NUT space.
Conversely, the modes of the $B$-field gauge transformation in the
isometric direction enters into the corresponding diffeomorphism by the
T-duality. This is a remnant of the mixing of the diffeomorphism and gauge
parameters \eqref{eq:NS5_zero-modes_parameters}.
Since the zero-modes of branes are determined by the local geometry,
even though the Q-brane shows the globally non-geometric nature, they
are still understandable in the sense of the ordinary space-time
picture.
In the effective action of the Q5-brane, the two geometric modes along
the isometry directions disappear in the pull-backs of the background fields while
the two modes in \eqref{eq:Q5_deltaB} show up as extra scalar fields \cite{Chatzistavrakidis:2013jqa,Kimura:2014upa,Blair:2017hhy}.
This is consistent with our result.

As in the cases of the KK-monopoles, the tensor zero-modes appear from
the higher ranks of the RR potentials whose indices extend along the isometry directions.

\paragraph{R5- and space-filling 5-branes}
The above picture of the zero-modes becomes obscure when we try to
understand locally non-geometric objects like the R-branes.
For the R5-brane, the solution loses the conventional geometric
meaning since it inevitably involves the winding coordinates.
There are two isometry directions in the R5-brane solution, namely, 
the $\bar{\Isy}_8$- and $\bar{\Isy}_9$-directions
in \eqref{eq:R5-brane}.
Corresponding to these directions, there are two independent zero-modes
associated with the $B$-field gauge transformations.
Since the R5-brane is a domain wall type object
in the conventional space-time, 
there is one geometric fluctuation mode along the ${\Isy}^6$-direction.
The last mode is given by the DFT gauge parameter $\xi_7 = \Lambda_7 = H^{-1} \phi^7$.
Since this parameter gives the translational symmetry along the
transverse direction to the R5-brane, 
it is natural to interpret it as the {\it fluctuation in the winding space}
rather than the $B$-field gauge transformation.
Since the R-brane dynamics is governed by the fluctuations along the
winding space, it should play a role of a probe for winding space.
This feature apparently distinguishes the R-brane from the conventional extended
objects in string theory.

The same is true even for the space-filling branes.
The solutions are 
localized in the winding space and spontaneously break the translational symmetry
along these directions.
Therefore, the two scalar fields on the space-filling brane \eqref{eq:sf-brane}
represent the fluctuations along the two winding directions.
The other two are zero-modes associated with the $B$-field gauge
transformations. If the two winding directions are smeared, the solution
becomes a flat space. In this case, all the four fluctuation zero-modes
correspond to the zero-modes by the $B$-field gauge transformations.

All the zero-modes of the non-geometric branes are summarized in Table
\ref{tb:Q5R5SF5_RRmodes1}. 
The corresponding gauge parameters in DFT are summarized in Table \ref{tab:fivebrane_gauge_parameter}.

\begin{table}[t] 
\centering
\begin{tabular}{l||c|c|c|c|c|c}
        & $\phi^6$ & $\phi^7$ & $\phi^8$ & $\phi^9$ & 0-form & 2-form 
\\
\hline
IIA Q5 & geometric & geometric & $\delta B$ & $\delta B$ & $\delta
		     C^{(3)} = {\dop} \lambda^{(2)} : \lambda^{(2)}_{89}$ &
			 $\delta C^{(5)} = {\dop} \lambda^{(4)} : \lambda^{(4)}_{mn89}$\\
\hline
IIB R5 & geometric & winding & $\delta B$ & $\delta B$ & $\delta C^{(4)} =
		     {\dop} \lambda^{(3)} : \lambda^{(3)}_{789}$ & $\delta
			 C^{(6)} = {\dop} \lambda^{(5)} :
			 \lambda^{(5)}_{mn789}$ 
\\
\hline
IIA SF5 & winding & winding & $\delta B$ & $\delta B$ & $\delta C^{(5)} =
		     {\dop} \lambda^{(4)} : \lambda^{(4)}_{6789}$ & $\delta
			 C^{(7)} = {\dop} \lambda^{(6)} :
			 \lambda^{(6)}_{mn6789}$ 
\\ \hline \hline
        & & & & & \multicolumn{2}{c}{1-form} 
\\
\hline
IIB Q5 & geometric & geometric & $\delta B$ & $\delta B$ & \multicolumn{2}{c}{$\delta
 C^{(4)} = {\dop} \lambda^{(3)} : \lambda^{(3)}_{m89}$} 
\\
\hline
IIA R5 & geometric & winding & $\delta B$ & $\delta B$ & \multicolumn{2}{c}{$\delta C^{(5)}
 = {\dop} \lambda^{(4)} : \lambda^{(4)}_{m789}$}
\\
 \hline
IIB SF5 & winding & winding & $\delta B$ & $\delta B$ & \multicolumn{2}{c}{$\delta C^{(6)}
 = {\dop} \lambda^{(5)} : \lambda^{(5)}_{m6789}$}
\end{tabular}
\caption{Physical interpretation of the zero-modes in the six-dimensional
 $\mathcal{N} = (2,0)$ tensor and the $\mathcal{N} = (1,1)$ vector
 multiplets. 
For the cases of the non-geometric Q5-, R5 and SF5-branes.
The index structure of the gauge parameters are specified explicitly.}
\label{tb:Q5R5SF5_RRmodes1}
\end{table}

\begin{table}[t]
\centering
\begin{tabular}{l | c c c c c}
 & NS5 & KKM & Q5 & R5 & SF5 \\
\hline
 diffeomorphism $\zeta^a$ & 
$ H^{-1} \phi^a $ & 
$\begin{pmatrix}
H^{-1} \phi^6 \\ H^{-1} \phi^7 \\ H^{-1} \phi^8 \\ 0
\end{pmatrix}$ & 
$\begin{pmatrix}
H^{-1} \phi^6 \\ H^{-1} \phi^7 \\ H^{-1} \phi^9 A_8 \\ 0
\end{pmatrix}$ & 
$\begin{pmatrix}
H^{-1} \phi^6 \\ 0 \\ H^{-1} \phi^9 A_8 \\ 0
\end{pmatrix}$ &
$\begin{pmatrix}
0 \\ 0 \\ H^{-1} \phi^9 A_8 \\ 0
\end{pmatrix}$ \\
gauge parameter $\Lambda_a$ & 
$ H^{-1} b_{ab} \phi^b $ & 
$\begin{pmatrix}
H^{-1} \phi^9 A_6 \\ H^{-1} \phi^9 A_7 \\ H^{-1} \phi^9 A_8 \\ H^{-1} \phi^9
\end{pmatrix}$ & 
$\begin{pmatrix}
0 \\ 0 \\ H^{-1} \phi^8 \\ H^{-1} \phi^9
\end{pmatrix}$ & 
$\begin{pmatrix}
0 \\ H^{-1} \phi^7 \\ H^{-1} \phi^8 \\ H^{-1} \phi^9
\end{pmatrix}$ &
$\begin{pmatrix}
H^{-1} \phi^6 \\ H^{-1} \phi^7 \\ H^{-1} \phi^8 \\ H^{-1} \phi^9
\end{pmatrix}$
\end{tabular}
\caption{The diffeomorphism and the $B$-field gauge parameters in each T-duality frame.}
\label{tab:fivebrane_gauge_parameter}
\end{table}

\subsection{Localized KK-monopole type}
Another type of the locally non-geometric object is the so-called
localized KK-monopole solution. 
This is first proposed as a genuine T-duality counterpart of the
NS5-brane \cite{Gregory:1997te}. Remember that when one relates the NS5-brane and the
KK-monopole by the T-duality transformation, one needs to introduce the
isometry along the transverse direction to the brane.
This is achieved by performing the smearing in the NS5-brane solution 
and the resulting geometry is known as the H-monopole of codimension
three. The isometry in the H-monopole corresponds to that of the
Taub-NUT space in the T-dualized KK-monopole side.
On the other hand, it was shown that the isometry in the H-monopole is
broken by the string worldsheet instanton effects \cite{Tong:2002rq}.
The original NS5-brane geometry is recovered by summing up all the
instanton effects which are interpreted as the KK-modes in the geometry.
The same phenomenon occurs in the KK-monopole side.
The isometry of the Taub-NUT space is again broken by the worldsheet
instanton effects \cite{Harvey:2005ab} and the geometry is modified due
to the T-dual of the KK-modes, namely, the string winding modes.
As a result, the modified geometry of the KK-monopole is characterized
by the winding coordinate $\tilde{x}$ and it ceases to be a solution to
supergravity. This is known as the localized KK-monopole solution.
Although it is not a solution to supergravity,  it is in fact a solution
to DFT \cite{Berman:2014jsa}.
This result carries over to the Q5-brane. 
The worldsheet instanton effects break the isometries in the Q5-brane
geometry and they introduce the winding coordinate dependence to the Q5-brane
geometry \cite{Kimura:2013fda, Kimura:2013zva}.
The modified Q5-brane is no longer a solution to supergravity but it
should be a solution to DFT. We call this the localized Q5-brane.
Indeed, these localized solutions are obtained by the formal T-duality
transformations of the NS5-brane geometry without isometries.
One notices that all the five-brane solutions discussed in Section
\ref{sct:five-branes_DFT} are written down by applying the formal T-duality
transformations to the NS5-brane irrespective of their codimensions.
Once a T-duality transformation is applied, one transverse geometric
coordinate (that corresponds to the Fourier conjugate of the KK-modes) is
switched to the winding coordinate corresponding to the Fourier conjugate of the
winding modes.
We therefore obtain the localized KK-monopole (KKw1:\ KK-monopole with one
winding coordinate dependence), the localized Q5-brane (Q5w2:\ Q5-brane
with two winding coordinate dependence), the localized R5-brane (R5w3)
and the localized SF5-brane (SF5w4). They are apparently locally
non-geometric objects and are indeed the solutions to DFT \cite{Kimura:2018hph}.

It is obvious that our calculations determining the zero-modes are
irrelevant to the explicit form of the harmonic function $H$. Only we
required is that it satisfies the Laplace equation $\Box H = 0$ in the
transverse directions. Therefore all the analysis discussed above are
also applied to the localized solutions. The effective theories of these
localized five-branes are governed by the six-dimensional 
$\mathcal{N} = (2,0)$ tensor and the $\mathcal{N} = (1,1)$ vector multiplets. 
The corresponding gauge parameters in DFT are summarized in Table \ref{tab:fivebrane_gauge_parameter_codim4}.
Again, a natural interpretation is that the modes given by $H^{-1}
\phi^a$ are fluctuations along the winding directions rather than the
$B$-field gauge transformations.

\begin{table}[t]
\centering
\begin{tabular}{l | c c c c c}
 & NS5 & KKw1 & Q5w2 & R5w3 & SF5w4 \\ \hline
 diffeomorphism $\zeta^a$ & 
$ H^{-1} \phi^a $ & 
$\begin{pmatrix}
H^{-1} \phi^6 \\ H^{-1} \phi^7 \\ H^{-1} \phi^8 \\ H^{-1} b_{9 b} \phi^b
\end{pmatrix}$ & 
$\begin{pmatrix}
H^{-1} \phi^6 \\ H^{-1} \phi^7 \\ H^{-1} b_{8b} \phi^b \\ H^{-1} b_{9b} \phi^b
\end{pmatrix}$ & 
$\begin{pmatrix}
H^{-1} \phi^6 \\ H^{-1} b_{7b} \phi^b \\ H^{-1} b_{8b} \phi^b \\ H^{-1} b_{9b} \phi^b
\end{pmatrix}$ &
$\begin{pmatrix}
H^{-1} b_{6b} \phi^b \\ H^{-1} b_{7b} \phi^b \\ H^{-1} b_{8b} \phi^b \\ H^{-1} b_{9b} \phi^b
\end{pmatrix}$ \\
gauge parameter $\Lambda_a$ & 
$H^{-1} b_{ab} \phi^b$ & 
$\begin{pmatrix}
H^{-1} b_{6b} \phi^b \\ H^{-1} b_{7b} \phi^b \\ H^{-1} b_{8b} \phi^b \\ H^{-1} \phi^9
\end{pmatrix}$ & 
$\begin{pmatrix}
H^{-1} b_{6b} \phi^b \\ H^{-1} b_{7b} \phi^b \\ H^{-1} \phi^8 \\ H^{-1} \phi^9
\end{pmatrix}$ & 
$\begin{pmatrix}
H^{-1} b_{6 b} \phi^b \\ H^{-1} \phi^7 \\ H^{-1} \phi^8 \\ H^{-1} \phi^9
\end{pmatrix}$ &
$\begin{pmatrix}
H^{-1} \phi^6 \\ H^{-1} \phi^7 \\ H^{-1} \phi^8 \\ H^{-1} \phi^9
\end{pmatrix}$
\end{tabular}
\caption{The diffeomorphism and the $B$-field gauge parameters in the localized five-branes}
\label{tab:fivebrane_gauge_parameter_codim4}
\end{table}

\section{Conclusion and discussions} \label{sct:conclusion}
In this paper, we studied the world-volume effective theories of
the five-branes in type II string theories within the formalism of double field theory (DFT).
These include the NS5-brane, the KK-monopole, the Q5-, R5-branes and the space-filling brane.
They appear in the T-duality orbits including the familiar NS5-branes.
Among other things, the R5- and the space-filling branes are kinds of the
locally non-geometric objects.
Although they are no longer solutions to conventional supergravity, they
are solutions to DFT.

We first determined the precise
zero-modes associated with the 
spontaneous breaking of the translational symmetry in the doubled space.
This symmetry is a part of the gauge symmetry in DFT given by the generalized Lie derivative.
We showed that there are four bosonic zero-modes that correspond to the
broken shift symmetry along the transverse directions to the five-branes
and 
they satisfy the Klein-Gordon equation in the six-dimensional world-volume.
For the five-branes that are solutions to supergravity, namely, for the
ordinary geometric
and the globally non-geometric branes, 
we demonstrated that parts of the scalar zero-modes are interpreted as transverse geometric fluctuation of
branes.
The remaining zero-modes come from the 
$B$-field gauge transformations.
Compared with these supergravity branes, the R5- and the space-filling
branes, 
have fluctuation zero-modes along the winding directions.
Indeed, this interpretation is necessary for the locally non-geometric
objects since their geometries are intrinsically characterized in the winding space.

We next determined the zero-modes associated with the gauge symmetry in
the RR sector.
They include the 1-form and the pair of the 0-form and the self-dual
2-form.
We find that these zero-modes are normalizable and they satisfy the Maxwell
equations for $p$-forms in the world-volume.
The four fluctuation scalar modes together with
these RR zero-modes are organized into the bosonic sectors of the six-dimensional
$\mathcal{N} = (1,1)$ vector and the $\mathcal{N} = (2,0)$ tensor
multiplets.

We also studied the other type of locally non-geometric objects in string theory.
They are branes of the so-called localized KK-monopole type.
When we discuss the genuine T-dualized object of the localized, the non-smeared NS5-brane (not the
H-monopole), it is inevitable to consider the localized KK-monopole in the winding space. 
The geometry of the localized KK-monopole is characterized by the
winding coordinate and it is interpreted as the string worldsheet
instanton effects.
The notion of the localized KK-monopole is generalized to those for the
Q-, R- and the space-filling branes. Their explicit solutions are
written down in the winding space.
These localized objects are not solutions to supergravity but are shown
to be solutions to DFT \cite{Kimura:2018hph}.
We determined the zero-modes of these localized solutions and find that
they are governed by the six-dimensional supermultiplets.
Although the solutions of the locally non-geometric five-branes are given in
the winding space, we note that their dynamics are governed by the
conventional field theories described by the six-dimensional supermultiplets.

We stress that the notion of the locally non-geometric objects in the
winding space is necessary for uncovering the nature of stringy
geometries. Indeed, they never give rise in conventional supergravity which is
based on the point particle (or the zero-slope limit of string) picture
of geometries. DFT is one of a useful formalism to capture stringy geometries with their
non-Riemannian structures \cite{Park:2013mpa ,Lee:2013hma,Berman:2019izh}.
A mathematically rigorous treatment of the doubled space would help us to
understand the nature of the winding geometries \cite{Mori:2019slw,
Ikeda:2020lxz, Mori:2020yih}.

We showed that the effective theories of the five-branes are given by
the bosonic sectors of the six-dimensional $\mathcal{N} = (2,0)$ tensor
and the $\mathcal{N} = (1,1)$ vector multiplets. 
Since the supergravity five-branes discussed in this paper are the half BPS
objects preserving 16 supercharges, we expect that the locally
non-geometric five-branes keep also the same supercharges.
It is interesting to determine the fermionic zero-modes associated with
the spontaneous breaking of supersymmetry in DFT
\cite{Hohm:2011nu,Jeon:2012hp}.
It is also interesting to study the other locally non-geometric branes.
For example, the U-duality generalization of DFT, known as exceptional
field theory (EFT), involves
many kinds of non-geometric objects 
\cite{Berman:2014hna, Blair:2014zba, Gunaydin:2016axc, Lust:2017bwq,
Bakhmatov:2017les}. 
We would come back to these issues in future studies.

\subsection*{Acknowledgments}
The work of K.S. is supported by Grant-in-Aid for JSPS Research Fellow,
JSPS KAKENHI Grant Number \texttt{JP20J13957}.
The work of S.S. is supported in part by Grant-in-Aid for Scientific
Research (C), JSPS KAKENHI Grant Number \texttt{JP20K03952}.

\begin{appendix}
\section{Quick introduction to double field theory} \label{sct:DFT}
We here introduce the brief summary of DFT.
More details are found in \cite{Hohm:2011dv,Hohm:2013bwa}.

\subsection{NSNS sector}
Fields and gauge symmetries in the NSNS sector are geometrically unified
in DFT.
DFT is defined in the doubled space in which T-duality is realized manifestly.
The doubled space is characterized by the coordinate given by 
\begin{align}
X^M = 
\begin{pmatrix}
\tilde{x}_\mu \\ x^\mu
\end{pmatrix},
\end{align}
where $x^\mu$ is the Fourier conjugate to the Kaluza-Klein (KK) modes of
a string, while $\tilde{x}_\mu$ is the conjugate to the string winding
modes.
The NSNS sector of type II supergravities is organized into an $O(D,D)$
tensor $\mathcal{H}_{MN}$ and a scalar $d$ known as the generalized
metric and the generalized dilaton, respectively.
We consider $D=10$ in the following.

These fields are parametrized as 
\begin{align}
{\mathcal H}_{MN} = 
\begin{pmatrix}
g^{\mu\nu} & - g^{\mu\rho} B_{\rho\nu} \\
B_{\mu\rho} g^{\rho\nu} & g_{\mu\nu} - B_{\mu\rho} g^{\rho\sigma} B_{\sigma\nu}
\end{pmatrix}, \qquad
e^{-2d} = \sqrt{-g} e^{-2\Phi}.
\end{align}
The index of the $O(D,D)$ tensor is raised and lowered by the $O(D,D)$ invariant metric
\begin{align}
\eta_{MN} =
\begin{pmatrix}
0 & \delta^\mu{}_\nu \\
\delta_\mu{}^\nu & 0
\end{pmatrix}, \qquad
\eta^{MN} = 
\begin{pmatrix}
0 & \delta_\mu{}^\nu \\
\delta^\mu{}_\nu & 0
\end{pmatrix}.
\end{align}
The action of the NSNS sector in DFT is given by
\begin{align}
S_{\text{DFT,NS}} = \int {\dop}^{2D} X \; e^{-2d} {\mathcal R} ({\mathcal H}, d),
\label{appeq:NSDFT_action}
\end{align}
where the generalized Ricci scalar ${\mathcal R}$ is defined by
\begin{align}
{\mathcal R} 
&= 4 \mathcal{H}^{MN} \partial_M \partial_N d - \partial_M \partial_N
 \mathcal{H}^{MN} - 4 \mathcal{H}^{MN} \partial_M d \partial_N d
+ 4 \partial_M \mathcal{H}^{MN} \partial_N d
\notag \\
& \quad
+ \frac{1}{8}
 \mathcal{H}^{MN} \partial_M \mathcal{H}^{KL} \partial_N
 \mathcal{H}_{KL}
 - \frac{1}{2} \mathcal{H}^{MN} \partial_M
 \mathcal{H}^{KL} \partial_K \mathcal{H}_{NL}.
\end{align}
This action is manifestly invariant under the $O(D,D)$ transformation.
All the quantities in DFT should satisfy the physical condition $\eta^{MN} \partial_M \partial_N * = 0$.
One can show that the DFT action \eqref{appeq:NSDFT_action} is invariant
under the following DFT gauge transformations:
\begin{align}
\delta_{\xi} \mathcal{H}_{MN} =& \ \widehat{\mathcal{L}}_{\xi} \mathcal{H}_{MN} = 
\xi^P \partial_P \mathcal{H}^{MN} + (\partial^M \xi_P - \partial_P
 \xi^M) \mathcal{H}^{PN} + (\partial^N \xi_P - \partial_P \xi^N) \mathcal{H}^{MP},
\notag \\
\delta_{\xi} d =& \ \widehat{\mathcal{L}}_{\xi} d = \xi^M \partial_M d -
 \frac{1}{2} \partial_M \xi^M,
\end{align}
provided that the strong constraint 
\begin{align}
\eta^{MN} \partial_M * \partial_N * = 0
\end{align}
is satisfied. 
One of the trivial solution to the strong constraint is that the derivative of the winding coordinates vanishes $\tilde{\partial}^\mu * = 0$. 
With the imposition of the condition $\tilde{\partial}^\mu * = 0$, the DFT action is reduced to the NSNS sector of type II supergravity:
\begin{align}
S_{\text{DFT,NS}} \xrightarrow{\tilde{\partial}^\mu * = 0} 
S_{\text{SUGRA,NS}} = 
\int {\dop}^D x \sqrt{-g} e^{-2\Phi} \left[
	R + 4 (\partial \Phi)^2 - {1 \over 12} H_{\mu\nu\rho} H^{\mu\nu\rho} \right].
\end{align}

\subsection{RR sector}
The RR sector of type II supergravities is described by an $O(D,D)$ Majorana
spinor in DFT \cite{Fukuma:1999jt, Hohm:2011zr, Hohm:2011dv}.
We introduce the gamma matrices $\Gamma^M$ satisfying the following Clifford algebra in $2D$ dimensions:
\begin{align}
\{ \Gamma^M, \Gamma^N \} = 2 \eta^{MN} \mathbf{1}.
\end{align}
The gamma matrices $\Gamma^M = (\Gamma_{\mu}, \Gamma^{\mu})$ are represented by
\begin{align}
\Gamma_{\mu} = \sqrt{2} \psi_{\mu}, \qquad 
\Gamma^{\mu} = \sqrt{2} \psi^{\mu},
\end{align}
where $\psi^{\mu}$ and $\psi_{\mu}$ are the fermionic creation and annihilation operators 
satisfying the following relations,
\begin{align}
\{
\psi_{\mu}, \psi^{\nu}
\} = \delta_{\mu} {}^{\nu},
\qquad 
\{
\psi_{\mu}, \psi_{\nu}
\} = 0,
\qquad 
\{
\psi^{\mu}, \psi^{\nu}
\} = 0,
\qquad 
(\psi_{\mu})^{\dagger} = \psi^{\mu}.
\end{align}
The Clifford vacuum $|0 \rangle$ is defined by 
\begin{align}
\psi_{\mu} |0 \rangle = 0, \qquad \langle 0 | 0 \rangle = 1.
\end{align}
A general $O(D,D)$ spinor is expanded by the basis $\psi^{\mu}$.
We consider a spinor $\chi$ whose component expansion is given by 
\begin{align}
\chi = \sum_{p=0}^D \frac{1}{p!} C_{\mu_1 \cdots \mu_p} \psi^{\mu_1} \cdots \psi^{\mu_p} |0 \rangle.
\end{align}
Here the coefficients $C_{\mu_1 \cdots \mu_p}$ are identified with the RR $p$-forms.

There is a group homomorphism $\rho : \mathrm{Pin}(D,D) \to O(D,D)$ such that 
$h = \rho (S)$ for $S \in \mathrm{Pin} (D,D)$, $h \in O(D,D)$.
This means we have the following relation
\begin{align}
S \Gamma_M S^{-1} = \Gamma_N h^N {}_M.
\end{align}
Therefore $S$ is a spinor representation of $h$ satisfying $h \eta h^T = \eta$.
Conversely, for any $h \in O(D,D)$, we have $\pm S \in \mathrm{Pin} (D,D)$.

One notices that a spinor in $\mathrm{Pin}(D,D)$ is decomposed into eigenstates $\chi_{\pm}$
associated with the eigenvalues $(-)^{N_F} = \pm 1$:
\begin{align}
(-)^{N_F} \chi_{\pm} = (\pm 1) \chi_{\pm}.
\end{align}
Here, $N_F = \sum_{\mu} \psi^{\mu} \psi_{\mu}$ is the fermion number operator.
Then, we have the following chiral spinors
\begin{align}
\chi_+ =& \ 
\left[
\frac{1}{0!} C^{(0)} + \frac{1}{2!} C^{(2)}_{\mu_1 \mu_2} \psi^{\mu_1} \psi^{\mu_2} + \cdots 
+ \frac{1}{10!} C^{(10)}_{\mu_1 \cdots \mu_{10}} \psi^{\mu_1} \cdots \psi^{\mu_{10}}
\right] |0 \rangle
\notag \\
\chi_- =& \ 
\left[
\frac{1}{1!} C^{(1)}_{\mu_1} \psi^{\mu_1} + \frac{1}{3!} C^{(3)}_{\mu_1 \cdots \mu_3} \psi^{\mu_1} \cdots \psi^{\mu_3}
+ \cdots + \frac{1}{9!} C^{(9)}_{\mu_1 \cdots \mu_9} \psi^{\mu_1} \cdots \psi^{\mu_9}
\right] |0 \rangle.
\end{align}
All the RR potentials in type IIA/IIB theories are incorporated in this formalism.
This is nothing but the democratic formulation of type II supergravity.
When we impose the chirality condition on the spinors, the 
$\mathrm{Pin} (D,D)$ symmetry is broken down to $\mathrm{Spin} (D,D)$ and we have either 
IIA or IIB theory.

The charge conjugation operator $C$ satisfying $C \Gamma^M C^{-1} = (\Gamma^M)^{\dagger}$
is defined by 
\begin{align}
C = (\psi^1 - \psi_1) (\psi^2 - \psi_2) \cdots (\psi^{10} - \psi_{10}).
\end{align}
This also satisfies the relations
\begin{align}
C \psi_{\mu} C^{-1} = \psi^{\mu}, \qquad 
C \psi^{\mu} C^{-1} = \psi_{\mu}.
\end{align}
A Dirac operator $\slashed{\del}$ is defined by 
\begin{align}
\slashed{\del} = \frac{1}{\sqrt{2}} \Gamma^M \del_M =
\psi^{\mu} \del_{\mu} + \psi_{\mu} \tilde{\del}^{\mu}.
\end{align}
This is an $O(D,D)$ invariant operator and nilpotent under the strong constraint:
\begin{align}
\slashed{\del}^2 = \frac{1}{4} \{ \Gamma^M, \Gamma^N \} \del_M \del_N = \frac{1}{2} \eta^{MN} \del_M \del_N = 0.
\end{align}
When we solve the strong constraint by imposing $\tilde{\del}^\mu * = 0$, the Dirac operator defines the field strengths of the RR potentials:
\begin{align}
\slashed{\del} \chi =& \sum_p \frac{1}{p!} \del_{\mu} C_{\mu_1 \cdots \mu_p} \psi^{\mu} \psi^{\mu_1} \cdots \psi^{\mu_p} |0 \rangle.
\end{align}

A spin representation $\mathbb{S} = \mathbb{S}^{\dagger} \in \mathrm{Spin}^{-} (10,10)$
of the generalized metric $\mathcal{H}$ is defined through the relation,
\begin{align}
\mathbb{S} \Gamma_M \mathbb{S}^{-1} = \Gamma_N \mathcal{H}^{N} {}_M.
\end{align}
Here $\mathcal{H}^N {}_M = \eta^{NP} \mathcal{H}_{PM}$ and $\mathcal{H}^t = \rho (\mathbb{S}^{\dagger}) = \mathcal{H}$.
Then the DFT action in the RR sector is given by \cite{Hohm:2011zr, Hohm:2011dv},
\begin{align}
S_{\text{DFT,RR}} = \int \! \dop^{2D} X \, 
\frac{1}{4} (\slashed{\del} \chi)^{\dagger} \mathbb{S} (\slashed{\del} \chi).
\end{align}
Since the RR sector is introduced in the democratic formulation, 
we impose the following self-duality constraint to reduce the
duplicated degrees of freedom:
\begin{align}
\slashed{\del} \chi = - \mathcal{K} \slashed{\del} \chi.
\end{align}
Here we have introduced $\mathcal{K} = C^{-1} \mathbb{S}$.
The energy-momentum tensor for the matter sector is defined by 
\begin{align}
\mathcal{E}^{MN} = - \frac{1}{16} \mathcal{H}^{(M} {}_P \overline{\slashed{\del} \chi} \Gamma^{N)P} \slashed{\del} \chi
\end{align}
Here $\bar{\chi} = \chi^{\dagger} C$ is the Dirac conjugation.

\section{Detailed calculations} \label{sct:detailed_calculations}
Here we exhibit the detailed calculations that are skipped in the main
text. 

\subsection{Calculations on the generalized Ricci tensor}
Remember that the generalized Ricci tensor is given by ${\mathcal
R}_{MN} = P_{MN}{}^{KL} {\mathcal K}_{KL}$, we first decompose the
${\mathcal K}$ tensor.
In the following, we solve the strong constraint by the condition $\bar{\partial}^{\mu} * = 0$.
By substituting the parametrization \eqref{eq:monopole(3.3)} 
of the generalized metric and dilaton, we find the following result:
\begin{align}
{\mathcal K}_{\mu\nu}
&= 2 \partial_\mu \nabla_\nu {\Isphi} 
- {\IsGamma}^\sigma_{\mu\nu} \nabla_\sigma {\Isphi} 
+ ({\IsB}_{(\nu|\kappa} {\Isg}^{\kappa\tau}) \Big(  
	{\Isg}^{\rho\sigma} {\IsH}_{|\mu)\tau\rho} \nabla_\sigma {\Isphi} 
	 \Big)
+ ({\IsB}_{\mu\tau} {\Isg}^{\tau\alpha}) ({\IsB}_{\nu\kappa} {\Isg}^{\kappa\beta}) \Big( 
	{\IsGamma}^\sigma_{\alpha\beta} \nabla_\sigma {\Isphi}
	\Big)
\notag \\
& \quad 
- {1 \over 4} \Big(
	{\Isg}^{\rho\tau} {\Isg}^{\sigma\kappa} {\IsH}_{(\mu|\sigma\rho} {\IsH}_{|\nu)\kappa\tau}
	+ 2 {\Isg}^{\rho\sigma} {\IsB}_{(\nu|\kappa} 
		({\IsH}_{|\mu)\tau\rho} + \partial_{|\mu)} {\IsB}_{\tau\rho}) 
		\partial_\sigma {\Isg}^{\tau\kappa}
	\Big)
\notag \\
& \quad
- {1 \over 2} ({\IsB}_{(\nu|\kappa} {\Isg}^{\kappa\tau}) \Big(  
	({\Isg}^{\rho\sigma} \partial_\sigma {\IsH}_{|\mu)\tau\rho} 
		+ {\IsH}_{|\mu)\tau\rho} \partial_\sigma {\Isg}^{\rho\sigma}
		+ {\Isg}^{\rho\sigma} {\IsH}_{|\mu)\tau\rho} {\IsGamma}_\sigma)
\notag \\
& \hspace{30mm}
	+ 2 {\Isg}^{\alpha\beta} {\IsGamma}^\sigma_{\alpha\tau} 
		\partial_\sigma {\IsB}_{|\mu)\beta}
	- \partial_\sigma {\IsB}_{\tau\rho} \partial_{|\mu)} {\Isg}^{\rho\sigma} 
	 \Big)
\notag \\
& \quad  
- \partial_\mu {\IsGamma}_\nu
+ {1 \over 2} \partial_\sigma {\IsGamma}^\sigma_{\mu\nu} 
+ {1 \over 2} {\IsGamma}^\sigma_{\mu\nu} {\IsGamma}_\sigma 
- {1 \over 2} {\Isg}_{(\nu|\tau} {\IsGamma}^\tau_{\rho\sigma} \partial_{|\mu)} {\Isg}^{\rho\sigma}
\notag \\
&\quad
- ({\IsB}_{(\nu|\kappa} {\Isg}^{\kappa\tau}) \Big(  
	{\IsB}_{|\mu)\beta} {\IsGamma}^\sigma_{\alpha\tau} \partial_\sigma {\Isg}^{\alpha\beta}
	\Big)
- {1 \over 2} ({\IsB}_{\mu\tau} {\Isg}^{\tau\alpha}) ({\IsB}_{\nu\kappa} {\Isg}^{\kappa\beta}) \Big( 
	\partial_\sigma {\IsGamma}^\sigma_{\alpha\beta} 
		+ {\IsGamma}^\sigma_{\alpha\beta} {\IsGamma}_\sigma
	\Big).
\end{align}
Here the ordinary Christoffel symbol for ${\Isg}$ is defined by ${\IsGamma}^\rho_{\mu\nu} = {1 \over 2}
{\Isg}^{\rho\sigma} (\partial_\mu {\Isg}_{\nu\sigma} + \partial_\nu
{\Isg}_{\mu\sigma} - \partial_\sigma {\Isg}_{\mu\nu})$.
Similarly, ${\mathcal K}_\mu{}^\nu$ is decomposed as 
\begin{align}
{\mathcal K}_\mu{}^\nu
&= {1 \over 2} \Big( 
	{\Isg}^{\tau\nu} {\Isg}^{\rho\sigma} {\IsH}_{\rho\mu\tau} \nabla_\sigma {\Isphi} \Big)
+ ({\IsB}_{\mu\tau} {\Isg}^{\tau\alpha}) \Big( 
	{\IsGamma}^\sigma_{\alpha\beta} {\Isg}^{\beta\nu} \nabla_\sigma {\Isphi} \Big)
\notag \\
& \quad 
- {1 \over 4} \Big( 
	{\Isg}^{\rho\sigma} ({\IsH}_{\rho\mu\tau} + \partial_\mu {\IsB}_{\tau\rho}) 
		\partial_\sigma {\Isg}^{\tau\nu}
	+ {\Isg}^{\tau\nu} ({\IsH}_{\rho\mu\tau} \partial_\sigma {\Isg}^{\rho\sigma}
		+ {\Isg}^{\rho\sigma} \partial_\sigma {\IsH}_{\rho\mu\tau}
		+ {\Isg}^{\rho\sigma} {\IsH}_{\rho\mu\tau} {\IsGamma}_\sigma) 
\notag \\
& \hspace{12mm}
	+ 2 {\Isg}^{\tau\alpha} {\IsGamma}^\sigma_{\alpha\beta} {\Isg}^{\beta\nu} 
		\partial_\sigma {\IsB}_{\mu\tau}
	- {\Isg}^{\nu\tau} \partial_{\mu} {\Isg}^{\rho\sigma} \partial_\sigma {\IsB}_{\tau\rho} \Big)
\notag \\
& \quad 
- {1 \over 2} \Big( 
	{\IsB}_{\mu\tau} {\IsGamma}^\sigma_{\alpha\beta} {\Isg}^{\beta\nu} 
		\partial_\sigma {\Isg}^{\tau\alpha} \Big)
- {1 \over 2} ({\IsB}_{\mu\tau} {\Isg}^{\tau\alpha}) \Big( 
	{\IsGamma}^\sigma_{\alpha\beta} \partial_\sigma {\Isg}^{\beta\nu}
	+ {\Isg}^{\beta\nu} \partial_\sigma {\IsGamma}^\sigma_{\alpha\beta} 
	+ {\IsGamma}^\sigma_{\alpha\beta} {\Isg}^{\beta\nu} {\IsGamma}_\sigma
	 \Big).
\end{align}
Finally, ${\mathcal K}^{\mu\nu}$ is decomposed as 
\begin{align}
{\mathcal K}^{\mu\nu}
&= {\Isg}^{\mu\tau} {\Isg}^{\nu\kappa} {\IsGamma}_{\tau\kappa}^\sigma \nabla_\sigma {\Isphi}
- {1 \over 2} {\Isg}^{\mu\tau} {\Isg}^{\nu\kappa} {\IsGamma}_{\tau\kappa}^\sigma 
	{\IsGamma}_\sigma
- {1 \over 2} \Big(
	2 {\Isg}^{(\mu|\tau} {\IsGamma}_{\tau\kappa}^\sigma \partial_\sigma {\Isg}^{|\nu)\kappa}
	+ {\Isg}^{\mu\tau} {\Isg}^{\nu\kappa} \partial_\sigma {\IsGamma}_{\tau\kappa}^\sigma \Big).
\end{align}
By using these expressions and the explicit form of the projector
$P_{MN} {}^{KL}$, the generalized Ricci tensor is written in the
following forms:
\begin{align}
{\mathcal R}_{\mu\nu}
&= {1 \over 2} (\text{${\Isg}$ eq.})_{\mu\nu}
	- ({\IsB}{\Isg}^{-1})_{(\mu|}{}^\beta (\text{${\IsB}$ eq.})_{|\nu)\beta}
	- {1 \over 2} ({\IsB}{\Isg}^{-1})_\mu{}^\alpha ({\IsB}{\Isg}^{-1})_\nu{}^\beta
		(\text{${\Isg}$ eq.})_{\alpha\beta}, 
\notag \\
{\mathcal R}_\mu{}^\nu
&= - {1 \over 2} {\Isg}^{\nu\beta} (\text{${\IsB}$ eq.})_{\mu\beta}
	- {1 \over 2} ({\IsB}{\Isg}^{-1})_\mu{}^\alpha {\Isg}^{\nu\beta}
		(\text{${\Isg}$ eq.})_{\alpha\beta},
\notag \\
{\mathcal R}^{\mu\nu}
&= - {1 \over 2} {\Isg}^{\mu\alpha} {\Isg}^{\nu\beta}
		(\text{${\Isg}$ eq.})_{\alpha\beta},
\end{align}
where we have defined the following expressions:
\begin{align}
(\text{${\Isg}$ eq.})_{\mu\nu}
&= \Italicsf{R}_{\mu\nu} - {1 \over 4} {\IsH}_{\mu\rho\sigma} {\IsH}_\nu{}^{\rho\sigma}
	+ 2 \nabla_\mu \nabla_\nu {\Isphi},
\notag \\
(\text{${\IsB}$ eq.})_{\mu\nu}
&= {1 \over 2} \nabla^\alpha H_{\alpha\mu\nu} 
	- H_{\alpha\mu\nu} \nabla^\alpha {\Isphi},
\notag \\
({\IsB}{\Isg}^{-1})_\mu{}^\nu
&= {\IsB}_{\mu\rho} {\Isg}^{\rho\nu}.
\end{align}

\subsection{Calculations on the RR zero-modes}
With the imposition of the strong constraint and by substituting the
ansatz \eqref{eq:RR_ansatz} with the promoted field 
$\lambda$ into the equation of motion \eqref{eq:RReom}, we
find the fluctuation satisfies 
\begin{align}
\left(
\psi^a \del_a + \psi^{m} \del_{m}
\right)
\left[
C 
\mathbb{S}_{\IsH}
\left(
\psi^b \del_b + \psi^{n} \del_{n}
\right)
\delta_{\lambda} \chi
\right] = 0,
\end{align}
where the indices $m,n, \ldots = 0,1, \ldots, 5$ and $a,b, \ldots = 6,
\ldots, 9$ run over the world-volume and the transverse directions.
Here ${\IsH}\,$ 
is the generalized metric associated with the
localized DFT monopole \eqref{eq:DFT_monopole_ansatz}.
We first evaluate the derivatives on the fluctuation in the bracket:
\begin{align}
\left(
\psi^b \del_b + \psi^{n} \del_{n}
\right) \delta_{\lambda} \chi
=& \ 
\frac{1}{2} e^{\frac{1}{2} b \psi \psi} 
\del_b b_{e'f'} \psi^{e'f'} 
\left[
\frac{1}{2} H^t \del_c b_{ef} \psi^{ef} + t H^{t-1} \del_c H
\right] 
\psi^{bc} \lambda
\notag \\
& \
+ e^{\frac{1}{2} b \psi \psi}
\left[
\frac{1}{2} t H^{t-1} \del_b H \del_c b_{ef} \psi^{ef} 
+ \frac{1}{2} H^t \del_b \del_c b_{ef} \psi^{ef}
\right.
\notag \\
& \qquad \qquad \qquad 
\left.
+ t (t-1) H^{t-1} \del_b H \del_c H 
+ t H^{t-1} \del_b \del_c H
\frac{}{}
\right]
\psi^{bc} \lambda
\notag \\
& \ 
- e^{\frac{1}{2} b \psi \psi} 
\left[
\frac{1}{2} H^t \del_c b_{ef} \psi^{ef} + t \del_c H
\right] \psi^c \psi^{n} \del_{n} \lambda.
\label{eq:Appcalc1}
\end{align}
Since the indices $a,b,c, \ldots$ run in the transverse directions to the five-brane, 
we have $\psi^{e'f'efbc} = 0$ due to the anti-symmetric nature of $\psi^a$.
Then the first term in the bracket in the first line vanishes.
By the factor $\psi^{bc}$, terms with the indices $bc$ are all
anti-symmetrized. This simplifies the expression \eqref{eq:Appcalc1} and
we have 
\begin{align}
\eqref{eq:Appcalc1} =& \ 
\frac{1}{2} t e^{\frac{1}{2} b \psi \psi} H^{t-1} \del_c H \del_b b_{ef} \psi^{efbc} \lambda
+
\frac{1}{2} t e^{\frac{1}{2} b \psi \psi} H^{t-1} \del_b H \del_c b_{ef} \psi^{efbc} \lambda
\notag \\
& \ 
- \frac{1}{2} e^{\frac{1}{2} b \psi \psi} H^s \del_c b_{ef} 
\sum_p \frac{1}{(p-1)!} \del_{n} \lambda_{m_1 \cdots m_{p-1}} 
\psi^{efc} \psi^{n m_1 \cdots m_{p-1}} | 0 \rangle
\notag \\
& \ 
- t e^{\frac{1}{2} b \psi \psi} H^{t-1} \del_c H 
\sum_p \frac{1}{(p-1)!} \del_{n} \lambda_{m_1 \cdots m_{p-1}} 
\psi^c \psi^{n m_1 \cdots m_{p-1}} |0 \rangle.
\end{align}
We next evaluate the term
\begin{align}
\mathbb{S}_{\IsH}
\left( 
\psi^b \del_b + \psi^{n} \del_{n}
\right) \delta_{\lambda} \chi.
\label{eq:appcalc2}
\end{align}
The spinor representation of the generalized metric is decomposed as 
\begin{align}
\mathbb{S}_{\IsH} = S_{\IsB}^\dag S_{\Isg}^{-1} S_{\IsB},
\qquad 
S_{\IsB} = e^{- \frac{1}{2} b_{ab} \psi^{ab}}, 
\qquad
S_{\IsB}^{\dagger} = e^{\frac{1}{2} b_{ab} \psi_a \psi_b}.
\end{align}
where $S_{\Isg} = S_{\Isg}^\dag$  
is the spinor representation of 
${\Isg}_{\mu\nu}$.
Using this expression we have 
\begin{align}
\eqref{eq:appcalc2} =& \ 
- \frac{1}{2} e^{\frac{1}{2} b_{ab} \psi_{ab}} S_{\Isg}^{-1} 
\left[
\frac{1}{2} H^t \del_c b_{ef} \sum_p \frac{1}{(p-1)!} \del_{n} \lambda_{m_1 \cdots m_{p-1}} 
\psi^{efc} \psi^{n m_1 \cdots m_{p-1}} |0 \rangle
\right]
\notag \\
& \ 
- t e^{\frac{1}{2} b_{ab} \psi_{ab}} S_{\Isg}^{-1} 
\left[
H^{t-1} \del_c H \sum_p \frac{1}{(p-1)!} \del_n \lambda_{m_1 \cdots m_{p-1}} 
\psi^c \psi^{n m_1 \cdots m_{p-1}} |0 \rangle
\right].
\label{eq:calc2-2}
\end{align}
Note that the unconventional index contractions with the annihilation
operator $\psi_{\mu}$.
By using the formula \cite{Hohm:2011dv}
\begin{align}
S_{\Isg}^{-1} \psi^{\mu_1} \cdots \psi^{\mu_p} | 0 \rangle 
= - \sqrt{|\det {\Isg}|} {\Isg}^{\mu_1 \nu_1} \cdots {\Isg}^{\mu_p \nu_p} \psi^{\nu_1} \cdots \psi^{\nu_p} |0 \rangle,
\label{eq:formula}
\end{align}
and the localized DFT monopole solution \eqref{eq:DFT_monopole_ansatz}, we find
\begin{align}
S_{\Isg}^{-1} \psi^{efc} \psi^{n m_1 \cdots m_{p-1}} |0 \rangle
=& \  - \sqrt{|\det {\Isg}|} {\Isg}^{ee'} {\Isg}^{ff'} {\Isg}^{cc'} {\Isg}^{nl} {\Isg}^{m_1 n_1} \cdots {\Isg}^{m_{p-1} n_{p-1}}
\psi^{e'f'c'} \psi^{l n_1 \cdots n_{p-1}} |0 \rangle
\notag \\
=& \ - \sqrt{H^4} H^{-1} \delta^{ee'} H^{-1} \delta^{ff'} H^{-1} \delta^{cc'}
\eta^{n l} \eta^{m_1 n_1} \cdots \eta^{m_{p-1} n_{p-1}}
\psi^{e'f'c'} \psi^{l n_1 \cdots n_{p-1}} |0 \rangle
\notag \\
=& \ 
- H^{-1} \eta^{n l} \eta^{m_1 n_1} \cdots \eta^{m_{p-1} n_{p-1}}
\psi^{efc} \psi^{l n_1 \cdots n_{p-1}} |0 \rangle.
\end{align}
Likewise, we have
\begin{align}
S_{\Isg}^{-1} \psi^c \psi^{n m_1 \cdots m_{p-1}} |0 \rangle
= - H \eta^{l n} \eta^{m_1 n_1} \cdots \eta^{m_{p-1} n_{p-1}}
\psi^c \psi^{l n_1 \cdots n_{p-1}} |0 \rangle.
\end{align}
Then,
\begin{align}
\eqref{eq:calc2-2} =& \ 
\frac{1}{2} e^{\frac{1}{2} b_{ab} \psi_{ab}} H^{t-1} 
\del_c b_{ef} \sum_p \frac{1}{(p-1)!} \del_{n} \lambda_{m_1 \cdots m_{p-1}} 
\eta^{n l} \eta^{m_1 n_1} \cdots \eta^{m_{p-1} n_{p-1}} 
\psi^{efc} \psi^{l n_1 \cdots n_{p-1}} | 0 \rangle
\notag \\
& \ 
+ t e^{\frac{1}{2} b_{ab} \psi_{ab}} H^t 
\del_c H \sum_p \frac{1}{(p-1)!} \del_{n} \lambda_{m_1 \cdots m_{p-1}} 
\eta^{n l} \eta^{m_1 n_1} \cdots \eta^{m_{p-1} n_{p-1}} 
\psi^c \psi^{l n_1 \cdots n_{p-1}} | 0 \rangle
\notag \\
=& \ 
\frac{1}{2} e^{\frac{1}{2} b_{ab} \psi_{ab}} H^{t-1} \del_c b_{ef} \sum_p f^{l n_1 \cdots n_{p-1}} 
\psi^{efc} \psi^{l n_1 \cdots n_{p-1}} | 0 \rangle
\notag \\
& \ 
+ t e^{\frac{1}{2} b_{ab} \psi_{ab}} H^t \del_c H \sum_p f^{l n_1 \cdots n_{p-1}} 
\psi^c \psi^{l n_1 \cdots n_{p-1}} |0 \rangle.
\end{align}
Here we have defined 
\begin{align}
f^{l n_1 \cdots n_{p-1}} = 
\frac{1}{(p-1)!} \del_{n} \lambda_{m_1 \cdots m_{p-1}} 
\eta^{n l} \eta^{m_1 n_1} \cdots \eta^{m_{p-1} n_{p-1}}.
\end{align}
Finally, we evaluate the quantity
\begin{align}
\left(
\psi^a \del_a + \psi^{m} \del_{m}
\right)
\left[
C 
\mathbb{S}_{\IsH}
\left(
\psi^b \del_b + \psi^{n} \del_{n}
\right)
\delta_{\lambda} \chi
\right].
\label{eq:appcalc3}
\end{align}
The charge conjugation operator $C$ satisfies $\psi^{\mu} C = C \psi_{\mu}$.
Then, we have 
\begin{align}
\eqref{eq:appcalc3} =& \ 
C 
\left(
\psi_a \del_a + \psi_{m} \del_{m}
\right)
\left[
\frac{1}{2} e^{\frac{1}{2} b_{ab} \psi_{ab}} H^{t-1} \del_c b_{ef}
\sum_p f^{l n_1 \cdots n_{p-1}} \psi^{efc} \psi^{l n_1 \cdots n_{p-1}} |0 \rangle
\right.
\notag \\
& \ 
\qquad \qquad \qquad \qquad \qquad 
\left.
+ t e^{\frac{1}{2} b_{ab} \psi_{ab}} H^t
\del_c H \sum_p f^{l n_1 \cdots n_{p-1}} 
\psi^c \psi^{l n_1 \cdots n_{p-1}} |0 \rangle
\right].
\label{eq:appcalc3-1}
\end{align}
Since all the annihilation operators are anti-commute with each other, we have
\begin{align}
\del_a e^{\frac{1}{2} b_{ab} \psi_{ab}} = \frac{1}{2} \del_a
 b_{ef} \psi_{ef} e^{\frac{1}{2} b_{ab} \psi_{ab}}
= \frac{1}{2} e^{\frac{1}{2} b_{ab} \psi_{ab}} \del_a b_{ef} \psi_{ef},
\end{align}
and then we obtain
\begin{align}
\eqref{eq:appcalc3-1} =& \ 
C \frac{1}{2} \psi_{a'} 
\Bigg[
e^{\frac{1}{2} b_{ab} \psi_{ab}} \frac{1}{2} \del_{a'} b_{e'f'} \psi_{e'f'} \del_c b_{ef} \cdot H^{t-1}
+ e^{\frac{1}{2} b_{ab} \psi_{ab}} \del_{a'} \del_c b_{ef} \cdot H^{t-1}
\notag \\
& \qquad \qquad 
+ e^{\frac{1}{2} b_{ab} \psi_{ab}} \del_c b_{ef} \cdot (s-1) H^{t-2} \del_{a'} H
\Bigg] \sum_p f^{l n_1 \cdots n_{p-1}} \psi^{efc} \psi^{l n_1 \cdots n_{p-1}} |0 \rangle
\notag \\
& \ 
+ C t \psi_{a'} 
\Bigg[
e^{\frac{1}{2} b_{ab} \psi_{ab}} \frac{1}{2} \del_{a'} b_{e'f'} 
\psi_{e'f'} H^t \del_c H
+ e^{\frac{1}{2} b_{ab} \psi_{ab}} \cdot t H^{t-1} \del_{a'} H \del_c H
\notag \\
& \ \qquad \qquad 
+ e^{\frac{1}{2} b_{ab} \psi_{ab}} H^t \del_{a'} \del_c H
\Bigg] \sum_p f^{l n_1 \cdots n_{p-1}} 
\psi^c \psi^{l n_1 \cdots n_{p-1}} |0 \rangle
\notag \\
& \ 
+ C \frac{1}{2} \psi_{m} 
\Bigg[
e^{\frac{1}{2} b_{ab} \psi_{ab}} H^{t-1} \del_c b_{ef} 
\Bigg]
\sum_p \del_{m} f^{l n_1 \cdots n_{p-1}} 
 \psi^{efc} \psi^{l n_1 \cdots n_{p-1}} |0 \rangle
\notag \\
& \ 
+ C t \psi_{m} 
\Bigg[
e^{\frac{1}{2} b_{ab} \psi_{ab}} H^t \del_c H 
\Bigg]
\sum_p \del_{m} f^{l n_1 \cdots n_{p-1}} 
 \psi^c \psi^{l n_1 \cdots n_{p-1}} |0 \rangle.
\label{eq:appcalc3-2}
\end{align}
When the number of the creation and the annihilation operators are not balanced, 
we have the vanishing contribution like
$ \psi_a \psi_{e'f'} \psi^c |0 \rangle = 0$.
Using this fact, we find
\begin{align}
\eqref{eq:appcalc3-2} =& \ 
C e^{\frac{1}{2} b_{ab} \psi_{ab}} 
\Bigg[
\frac{1}{4} \del_{a'} b_{e'f'} \del_c b_{ef} H^{t-1} \psi_{a'e'f'} \psi^{efc}
+
\frac{1}{2} \del_{a'} \del_c b_{ef} \cdot H^{t-1} \psi_{a'} \psi^{efc}
\notag \\
& \
\qquad \qquad \quad
+
\frac{1}{2} (t-1) \del_c b_{ef} H^{t-2} \del_{a'} H \psi_{a'} \psi^{efc}
\notag \\
& \
\qquad \qquad \quad
+
t^2 H^{t-1} \del_{a'} H \del_c H \psi_{a'} \psi^c
+
t H^t \del_{a'} \del_c H \psi_{a'} \psi^c
\Bigg] \sum_p f^{l n_1 \cdots n_{p-1}} \psi^{l n_1 \cdots n_{n-p}} |0 \rangle
\notag \\
& \ 
+ C e^{\frac{1}{2} b_{ab} \psi_{ab}} 
\Bigg[
- \frac{1}{2} H^{t-1} \del_c b_{ef} \psi^{efc} 
- t H^{t} \del_c H \psi^c 
\Bigg] 
\sum_p \del_{m} f^{l n_1 \cdots n_{p-1}}
\psi_{m} \psi^{l n_1 \cdots n_{p-1}} |0 \rangle.
\label{eq:calc3-3}
\end{align}
The equation for zero-modes is therefore $\eqref{eq:calc3-3} = 0$.
Since $\psi_a$, $\psi^c$ anti-commutes with $\psi^{m}$ and 
\begin{align}
\psi_a \psi^c |0 \rangle = 
(\delta_a {}^c - \psi^c \psi_a ) | 0 \rangle 
= \delta_a {}^c |0 \rangle,
\end{align}
we have 
\begin{align}
t H^t \del_a \del_c H \psi_a \psi^c |0 \rangle
=
t H^t \Box H |0 \rangle = 0.
\end{align}
Here we have used the fact that $H$ is a harmonic function.
On the other hand, by using $\{\psi^a, \psi_b\} = \delta^a {}_b$ iteratively, we find
\begin{align}
\psi_a \psi^{efc} |0 \rangle 
= (\delta_a {}^e \psi^{fc} - \delta_a {}^f \psi^{ec} + \delta_a {}^c
 \psi^{ef}) |0 \rangle.
\end{align}
Then we find 
\begin{align}
\frac{1}{2} \del_a \del_c b_{ef} \cdot H^{t-1} \psi_a \psi^{efc} |0\rangle
=
\frac{1}{2} H^{t-1} 
\left(
\del_a \del_c b_{ab} + \del_a \del_b b_{ca} + \Box b_{bc}
\right) \psi^{bc} |0 \rangle.
\end{align}
From the BPS condition for the DFT monopole \eqref{eq:BPS_condition}, we
have the relation,
\begin{align}
\left(
\Box b_{bc} + \del_a \del_b b_{ca} + \del_a \del_c b_{ab}
\right) = \varepsilon_{abcd} \del_a \del_d H = 0.
\end{align}
Similarly, we have
\begin{align}
& \quad \
\frac{1}{2} (t-1) \del_c b_{ef} H^{t-2} \del_a H \psi_a \psi^{efc} |0
 \rangle
\notag \\
&= \frac{1}{2} (t-1) H^{t-2} \del_{[c} b_{ef]} \del_a H \psi_a \psi^{efc} |0 \rangle
\notag \\
&
= \frac{1}{2} (t-1) H^{t-2} 
\frac{1}{3} \varepsilon_{cefd} \del_d H \del_a H 
\left(
\delta_a {}^e \psi^{fc} - \delta_a {}^f \psi^{ec} + \delta_a {}^c \psi^{ef}
\right) |0 \rangle
\notag \\
& = 0.
\end{align}
We have also 
\begin{align}
t^2 H^{t-1} \del_a H \del_c \psi_a \psi^c |0 \rangle
=
t^2 H^{t-1} \del_a H \del_a H |0 \rangle
\end{align}
Finally, by using the relation
\begin{align}
\psi_{abc} \psi^{def} |0 \rangle
=& \ 
- 
\left(
\delta_a {}^d \delta_b {}^e \delta_c {}^f
-
\delta_a {}^d \delta_b {}^f \delta_c {}^e
+
\delta_a {}^e \delta_b {}^f \delta_c {}^d
-
\delta_a {}^e \delta_b {}^d \delta_c {}^f
+
\delta_a {}^f \delta_b {}^d \delta_c {}^e
-
\delta_a {}^f \delta_b {}^e \delta_c {}^d
\right) |0 \rangle
\notag \\
=& \ - 3! \delta_{a} {}^{[d} \delta_b {}^{e} \delta_{c} {}^{f]} |0 \rangle
\end{align}
we find 
\begin{align}
\frac{1}{4} H^{t-1}
\del_a b_{bc} \del_d b_{ef} \psi_{abc} \psi^{def} |0 \rangle
= - H^{t-1} \del_a H \del_a H |0 \rangle.
\end{align}
Therefore, we find that the first and the second lines in \eqref{eq:calc3-3} gives
\begin{align}
\Bigg[
- H^{t-1} \del_a H \del_a H 
+ t^2 H^{t-1} \del_a H \del_a H
\Bigg] \sum_{p} f^{l n_1 \cdots n_{p-1}}
\psi^{l n_1 \cdots n_{p-1}} |0 \rangle.
\end{align}
From the expression in \eqref{eq:calc3-3}, this terms should vanish by itself.
Therefore, we find,
\begin{align}
t = \pm 1.
\end{align}

On the other hand, by using $\{\psi_i, \psi^j \} = \delta_i {}^j$ several times, we find 
\begin{align}
\psi_{m} \psi^{l n_1 \cdots n_{p-1}} |0 \rangle 
= 
\left(
\delta_{m} {}^{l} \psi^{n_1 \cdots n_{p-1}} 
-
\delta_{m} {}^{n_1} \psi^{l n_2 \cdots
 n_{p-1}}
+ \cdots 
+ (-)^{p-1} \delta_{m} {}^{n_{p-1}} \psi^{l n_1
 \cdots n_{p-2}}
\right) | 0 \rangle
\end{align}
Then we find that the terms in the second bracket in \eqref{eq:calc3-3} vanishes when 
\begin{align}
\del_{m_1} F^{m_1 \cdots m_p} = 0,
\label{eq:maxwell}
\end{align}
where we have defined
\begin{align}
F^{m_1 m_2 \cdots m_p} =
p! \del^{[m_1} \lambda^{m_2 \cdots m_p]}.
\end{align}
The condition \eqref{eq:maxwell} is nothing but the Maxwell equation for the $p$-forms.

\subsection{Calculations on the normalizability}
We here show the derivation of the equation \eqref{eq:effective_RR}.
The derivative of the fluctuation results in 
\begin{align}
\slashed{\del} \delta \chi =& \ 
- \frac{1}{2} e^{\frac{1}{2} b_{ab} \psi^{ab}} 
H^t \del_c b_{ef} \sum_p f_{n m_1 \cdots m_{p-1}}
\psi^{efc} \psi^{n m_1 \cdots m_{p-1}} |0 \rangle
\notag \\
& \ 
-t e^{\frac{1}{2} b_{ab} \psi^{ab}} H^{t-1} \del_c H \sum_p f_{n
 m_1 \cdots m_{p-1}}
\psi^c \psi^{n m_1 \cdots m_{p-1}} |0 \rangle.
\end{align}
By using the relation 
$
\left(\psi^{\mu_1} \cdots \psi^{\mu_p} |0 \rangle \right)^{\dagger}
=
\langle 0 | \psi_{\mu_p} \cdots \psi_{\mu_1} 
$
and the fact 
$
(e^{\frac{1}{2} b_{ab} \psi^{ab}})^{\dagger} = e^{-
 \frac{1}{2} b_{ab} \psi_{ab}}
$,
we obtain
\begin{align}
(\slashed{\del} \delta \chi)^{\dagger}
=& \  - \frac{1}{2} H^t \del_c b_{ef} \sum_p f_{n m_1 \cdots
 m_{p-1}}
\langle 0 | \psi_{m_{p-1} \cdots m_{1} n} \psi_{cfe}
 e^{- \frac{1}{2} b_{ab} \psi_{ab}}
\notag \\
& \ 
- t H^{t-1} \del_c H \sum_p f_{n m_1 \cdots m_{p-1}} 
\langle 0 | \psi_{m_{p-1} \cdots m_{1} n} \psi_{c}
 e^{- \frac{1}{2} b_{ab} \psi_{ab}}.
\end{align}
Further, we have
\begin{align}
\mathbb{S}_{\IsH}\,
\slashed{\del} \delta \chi 
=& \ 
\frac{1}{2} e^{\frac{1}{2} b_{ab} \psi_{ab}} H^{t-1} \del_c b_{ef} 
\sum_p f_{n \cdots m_{p-1}} 
\eta^{n l} \cdots \eta^{m_{p-1} n_{p-1}}
\psi^{efc} \psi^{l \cdots n_{p-1}} |0 \rangle
\notag \\
& \ 
+ t e^{\frac{1}{2} b_{ab} \psi_{ab}} H^t \del_c H 
\sum_p f_{n \cdots m_{p-1}} 
\eta^{n l} \cdots \eta^{m_{p-1} n_{p-1}}
\psi^{c} \psi^{l \cdots n_{p-1}} |0 \rangle.
\end{align}
Then we find
\begin{align}
(\slashed{\del} \delta \chi)^{\dagger} 
\mathbb{S}_{\IsH}\, 
 (\slashed{\del} \delta \chi) 
=& \ 
- \frac{1}{4} H^{2t-1} \del_c b_{ef} \del_{c'} b_{e'f'} \sum_{p}
 f_{n m_1 \cdots m_{p-1}} \sum_{q} f_{n'
 m'_1 \cdots m'_{q-1}}
\eta^{n'l'} \eta^{m'_1 n'_1} \cdots
 \eta^{m'_{q-1} n'_{q-1}} 
\notag \\
& \ 
\qquad \qquad 
\times 
\langle 0 | \psi_{m_{p-1} \cdots m_1} \psi_{cfe} 
\psi^{e'f'c'} \psi^{l' n'_1 \cdots n'_{q-1}} |0
 \rangle
\notag \\
& \ 
- t^2 H^{2t-1} \del_c H \del_{c'} H 
\sum_{p}
 f_{n m_1 \cdots m_{p-1}} \sum_{q} f_{n'
 m'_1 \cdots m'_{q-1}}
\eta^{n'l'} \eta^{m'_1 n'_1} \cdots
 \eta^{m'_{q-1} n'_{q-1}} 
\notag \\
& \ 
\qquad \qquad 
\times
\langle 0 | \psi_{m_{p-1} \cdots m_1} \psi_{c} 
\psi^{c'} \psi^{l' n'_1 \cdots n'_{q-1}} |0
 \rangle.
\end{align}

\subsection{Calculations on the self-duality constraint}
Here we show the derivation of the conditions 
\eqref{eq:sd_comp2}.
The self-duality constraint of the $O(D,D)$ spinor fluctuation $\delta_\lambda \chi$ is given by 
\begin{align}
\slashed{\del} \delta_{\lambda} \chi = C 
\mathbb{S}_{\IsH}\, 
(\slashed{\del} \delta_{\lambda} \chi).
\label{eq:appsd2}
\end{align}
By using the relation $C \psi_\mu = \psi^\mu C$, 
$C S_{\Isg}^{-1} = - S_{\Isg} C$,
we find 
\begin{align}
C \mathbb{S}_{\IsH} 
=
C e^{\frac{1}{2} b_{ab} \psi_{ab}} S_{\Isg}^{-1} e^{- \frac{1}{2} b_{cd} \psi^{cd}}
= 
- e^{\frac{1}{2} b_{ab} \psi^{ab}} S_{\Isg} C e^{- \frac{1}{2} b_{cd} \psi^{cd}}.
\end{align}
Then the equation \eqref{eq:appsd2} becomes 
\begin{align}
\slashed{\del} \delta_{\lambda} \chi = 
- e^{\frac{1}{2} b_{ab} \psi^{ab}} S_{\Isg} C e^{- \frac{1}{2} b_{cd} \psi^{cd}} \slashed{\del} \delta_{\lambda} \chi.
\label{eq:appsd3}
\end{align}
On the other hand, using the calculations in the previous subsection, the left-hand side in \eqref{eq:appsd3} becomes
\begin{align}
\slashed{\del} \delta_{\lambda} \chi =& \ 
- \frac{1}{6} e^{\frac{1}{2} b_{ab} \psi^{ab}} H^t \del_d H \varepsilon_{cefd} \sum_{p} F_{n m_1 \cdots m_{p-1}} 
\psi^{cef} \psi^{n m_1 \cdots m_{p-1}} |0 \rangle
\notag \\
& \ 
- t e^{\frac{1}{2} b_{ab} \psi^{ab}} H^{t-1} \del_c H 
\sum_p F_{n m_1 \cdots m_{p-1}} \psi^c \psi^{n m_1 \cdots m_{p-1}} |0 \rangle.
\end{align}
Here we have used $\del_{[a} b_{bc]} = \frac{1}{3} \varepsilon_{abcd} \del_d H$ and introduced 
$
F_{\mu_1 \mu_2 \cdots \mu_p} = p! \del_{[\mu_1} \lambda_{\mu_2 \cdots
\mu_p]}
$.
The right-hand side in \eqref{eq:appsd3} is calculated as 
\begin{align}
- e^{\frac{1}{2} b_{ab} \psi^{ab}} S_{\Isg} C e^{- \frac{1}{2} b_{cd} \psi^{cd}} \slashed{\del} \delta_{\lambda} \chi
=& \ 
\frac{1}{6} e^{\frac{1}{2} b_{ab} \psi^{ab}} H^t \del_d H \varepsilon_{cefd} \sum_{p=1}^6 
F_{n m_1 \cdots m_{p-1}} 
S_{\Isg} C \psi^{cef} \psi^{n m_1 \cdots m_{p-1}} |0 \rangle
\notag \\
& \ 
+ t e^{\frac{1}{2} b_{ab} \psi^{ab}} H^{t-1} \del_c H \sum_{p=1}^6 F_{n m_1 \cdots m_{p-1}} 
S_{\Isg} C \psi^c \psi^{n m_1 \cdots m_{p-1}} |0 \rangle.
\label{eq:appsd_e}
\end{align}
We now evaluate the first and the second lines in \eqref{eq:appsd_e} separately.
For the first line, since $C \psi^\mu = \psi_\mu C$, we have 
\begin{align}
S_{\Isg} C \psi^{cef} \psi^{n m_1 \cdots m_{p-1}} |0 \rangle
= S_{\Isg} \psi_{cef} \psi_{n m_1 \cdots m_{p-1}} C |0 \rangle.
\label{eq:appsd3-1}
\end{align}
Further, since 
\begin{align}
C | 0 \rangle = (\psi^0 + \psi_0) (\psi^1 + \psi_1) \cdots (\psi^9 + \psi_9) | 0 \rangle
= \psi^{012 \cdots 9} |0 \rangle,
\end{align}
by decomposing the indices into the world-volume (012345) and the
transverse (6789) directions, we find
\begin{align}
\eqref{eq:appsd3-1}
= S_{\Isg} \psi_{cef} \psi^{6789} \psi_{n m_1 \cdots m_{p-1}} \psi^{012345} |0 \rangle.
\label{eq:appsd3-2}
\end{align}
Then using the formulae, 
\begin{align}
\psi_{cef} \psi^{6789} |0 \rangle 
=& \ 
\frac{1}{(4-3)!} \varepsilon_{fech} \psi^h | 0 \rangle,
\notag \\
\psi_{n m_1 \cdots m_{p-1}} \psi^{012345} |0 \rangle 
=& \
 \frac{1}{(6-p)!} 
\varepsilon_{
n m_1 \cdots m_{6-p} k_1 \cdots k_{6-p}
} \psi^{k_1 \cdots k_{6-p}} |0 \rangle,
\end{align}
we obtain 
\begin{align}
\eqref{eq:appsd3-2} =& \ 
\frac{1}{(6-p)!} \varepsilon_{fech} 
\varepsilon_{
n m_1 \cdots m_{6-p} k_1 \cdots k_{6-p}
} 
S_g \psi^h
\psi^{k_1 \cdots k_{6-p}} |0 \rangle.
\label{eq:appsd3-3}
\end{align}
Finally, by using the formula 
\begin{align}
S_{\Isg} \psi^{\mu_1 \cdots \mu_p} |0 \rangle 
= - \frac{1}{\sqrt{|\det {\Isg}|}} {\Isg}_{\mu_1 \nu_1} \cdots {\Isg}_{\mu_p \nu_p} 
\psi^{\nu_1 \cdots \nu_p} |0 \rangle
\end{align}
and substituting the explicit solution of the localized DFT monopole, we find 
\begin{align}
\eqref{eq:appsd3-3} =& \ 
\frac{1}{(6-p)!} 
\varepsilon_{fech} 
\varepsilon_{
n m_1 \cdots m_{6-p} k_1 \cdots k_{6-p}
} 
(- H^{-2}) H \delta^{hd} 
\eta_{k_1 l_1} \cdots \eta_{k_{6-p} l_{6-p}}
\psi^d \psi^{l_1 \cdots l_{6-p}}
 |0 \rangle.
\label{eq:appsd3-4}
\end{align}
Likewise, for the second line in \eqref{eq:appsd_e}, we have 
\begin{align}
S_{\Isg} 
C \psi^c \psi^{n m_1 \cdots m_{p-1}} |0 \rangle
= - \frac{1}{3!} \frac{1}{(6-p)!} H 
 \varepsilon_{cefh} 
\varepsilon_{n m_1 \cdots m_{p-1} l_1 \cdots l_{6-p}} 
\psi^{efh} \psi^{l_1 \cdots l_{6-p}} |0 \rangle.
\end{align}
Therefore, 
\begin{align}
&
\text{RHS of } \eqref{eq:appsd3} 
\notag \\
=& \ 
e^{\frac{1}{2} b_{ab} \psi^{ab}}
\Bigg[
- \frac{1}{6} \frac{1}{(6-p)!} H^{t-1} \del_d H 
\sum_{p} \varepsilon_{n m_1 \cdots m_{p-1} k_1 \cdots k_{6-p}} 
\eta_{k_1 l_1} \cdots \eta_{k_{6-p} l_{6-p}} 
F_{n m_1 \cdots m_{p-1}}
\psi^{d'} \psi^{l_1 \cdots l_{6-p}} |0 \rangle
\notag \\
& \ 
- t \frac{1}{3!} \frac{1}{(6-p)!} H^t \del_c H \varepsilon_{cefh}
\sum_p \varepsilon_{n m_1 \cdots m_{p-1} k_1 \cdots k_{6-p}}
\eta_{k_1 l_1} \cdots \eta_{k_{6-p} l_{6-p}}
F_{n m_1 \cdots m_{p-1}} 
\psi^{efh} \psi^{l_1 \cdots l_{p-1}} |0 \rangle
\Bigg]
\end{align}
On the other hand, we evaluate
\begin{align}
\text{LHS of } \eqref{eq:appsd3} =& \ 
e^{\frac{1}{2} b_{ab} \psi^{ab}}
\Bigg[
- t H^{t-1} \del_c H \sum_p F_{n m_1 \cdots m_{p-1}} \psi^c \psi^{n m_1 \cdots m_{p-1}} |0 \rangle
\notag \\
& \qquad \qquad \quad 
- \frac{1}{6} H^t \del_d H \varepsilon_{cefd} \sum_p F_{n m_1 \cdots m_{p-1}} 
\psi^{cef} \psi^{n m_1 \cdots m_{p-1}} |0 \rangle
\Bigg]
\end{align}
Then, by comparing the both sides of \eqref{eq:appsd3}, we obtain the
condition
\begin{align}
&
\sum_{p=0}^6 \frac{1}{(6-p)!} \varepsilon_{n m_1 \cdots m_{p-1} k_1 \cdots k_{6-p}}
\eta_{k_1 l_1} \cdots \eta_{k_{6-p} l_{6-p}} F_{n m_1 \cdots m_{p-1}}
\psi^{l_1 \cdots l_{6-p}}
=
- t \sum_{p=0}^6 F_{l_1 \cdots l_p} \psi^{l_1 \cdots l_{p}},
\notag 
\\
&
\sum_{p=0}^6 \frac{1}{(6-p)!} \varepsilon_{n m_1 \cdots m_{p-1} k_1 \cdots k_{6-p}}
\eta_{k_1 l_1} \cdots \eta_{k_{6-p} l_{6-p}} F_{n m_1 \cdots m_{p-1}}
\psi^{l_1 \cdots l_{6-p}}
=
- \frac{1}{t} \sum_{p=0}^6 F_{l_1 \cdots l_p} \psi^{l_1 \cdots l_{p}}.
\label{eq:appsd_comp2}
\end{align}
They are the conditions  
\eqref{eq:sd_comp2}.

\end{appendix}



\begin{thebibliography}{00}

\bibitem{Hull:1994ys}
C.~Hull and P.~Townsend,
``Unity of superstring dualities,''
Nucl. Phys. B \textbf{438} (1995), 109-137
[arXiv:hep-th/9410167 [hep-th]].

\bibitem{Blau:1997du}
M.~Blau and M.~O'Loughlin,
``Aspects of U duality in matrix theory,''
Nucl. Phys. B \textbf{525} (1998), 182-214
[arXiv:hep-th/9712047 [hep-th]].

\bibitem{Obers:1998fb}
  N.~A.~Obers and B.~Pioline,
  ``U-duality and M-theory,''
  Phys.\ Rept.\  {\bf 318} (1999) 113
  [hep-th/9809039].


\bibitem{Eyras:1999at}
E.~Eyras and Y.~Lozano,
``Exotic branes and nonperturbative seven-branes,''
Nucl. Phys. B \textbf{573} (2000), 735-767
[arXiv:hep-th/9908094 [hep-th]].

\bibitem{deBoer:2010ud}
  J.~de Boer and M.~Shigemori,
  ``Exotic branes and non-geometric backgrounds,''
  Phys.\ Rev.\ Lett.\  {\bf 104} (2010) 251603
  [arXiv:1004.2521 [hep-th]],
  ``Exotic branes in string theory,''
  Phys.\ Rept.\  {\bf 532} (2013) 65
  [arXiv:1209.6056 [hep-th]].


\bibitem{Fernandez-Melgarejo:2018yxq}
J.~J.~Fern{\'a}ndez-Melgarejo, T.~Kimura and Y.~Sakatani,
``Weaving the Exotic Web,''
JHEP \textbf{09} (2018), 072
[arXiv:1805.12117 [hep-th]].

\bibitem{Berman:2018okd}
D.~S.~Berman, E.~T.~Musaev and R.~Otsuki,
``Exotic Branes in Exceptional Field Theory: $E_{7(7)}$ and Beyond,''
JHEP \textbf{12} (2018), 053
[arXiv:1806.00430 [hep-th]].


\bibitem{Kimura:2013khz}
T.~Kimura and S.~Sasaki,
``Worldsheet Description of Exotic Five-brane with Two Gauged Isometries,''
JHEP \textbf{03} (2014), 128
[arXiv:1310.6163 [hep-th]].


\bibitem{Kimura:2014bea}
T.~Kimura, S.~Sasaki and M.~Yata,
``Hyper-Kahler with torsion, T-duality, and defect $(p, q)$ five-branes,''
JHEP \textbf{03} (2015), 076
[arXiv:1411.3457 [hep-th]].


\bibitem{Kikuchi:2012za}
T.~Kikuchi, T.~Okada and Y.~Sakatani,
``Rotating string in doubled geometry with generalized isometries,''
Phys. Rev. D \textbf{86} (2012), 046001
[arXiv:1205.5549 [hep-th]].

\bibitem{Okada:2014wma}
T.~Okada and Y.~Sakatani,
``Defect branes as Alice strings,''
JHEP \textbf{03} (2015), 131
[arXiv:1411.1043 [hep-th]].

\bibitem{Sakatani:2014hba}
Y.~Sakatani,
``Exotic branes and non-geometric fluxes,''
JHEP \textbf{03} (2015), 135
[arXiv:1412.8769 [hep-th]].


\bibitem{Hassler:2013wsa}
  F.~Ha{\ss}ler and D.~L\"{u}st,
  ``Non-commutative/non-associative IIA (IIB) $Q$- and $R$-branes and their intersections,''
  JHEP {\bf 1307} (2013) 048
  [arXiv:1303.1413 [hep-th]].


\bibitem{Hull:2004in}
C.~Hull,
``A Geometry for non-geometric string backgrounds,''
JHEP \textbf{10} (2005), 065
[arXiv:hep-th/0406102 [hep-th]].


\bibitem{Hull:2009mi}
  C.~Hull and B.~Zwiebach,
  ``Double Field Theory,''
  JHEP {\bf 0909} (2009) 099
  [arXiv:0904.4664 [hep-th]].

\bibitem{Siegel:1993xq}
  W.~Siegel,
  ``Two vierbein formalism for string inspired axionic gravity,''
  Phys.\ Rev.\ D {\bf 47} (1993) 5453
  [hep-th/9302036],
  ``Superspace duality in low-energy superstrings,''
  Phys.\ Rev.\ D {\bf 48} (1993) 2826
  [hep-th/9305073],
  ``Manifest duality in low-energy superstrings,''
  hep-th/9308133.


\bibitem{Berkeley:2014nza}
  J.~Berkeley, D.~S.~Berman and F.~J.~Rudolph,
  ``Strings and branes are waves,''
  JHEP {\bf 1406} (2014) 006
  [arXiv:1403.7198 [hep-th]].

\bibitem{Berman:2014jsa}
  D.~S.~Berman and F.~J.~Rudolph,
  ``Branes are waves and monopoles,''
  JHEP {\bf 1505} (2015) 015
  [arXiv:1409.6314 [hep-th]].

\bibitem{Bakhmatov:2016kfn}
  I.~Bakhmatov, A.~Kleinschmidt and E.~T.~Musaev,
  ``Non-geometric branes are DFT monopoles,''
  JHEP {\bf 1610} (2016) 076
  [arXiv:1607.05450 [hep-th]].

\bibitem{Kimura:2018hph}
T.~Kimura, S.~Sasaki and K.~Shiozawa,
``Worldsheet Instanton Corrections to Five-branes and Waves in Double Field Theory,''
JHEP \textbf{07} (2018), 001
[arXiv:1803.11087 [hep-th]].



\bibitem{Plauschinn:2018wbo}
E.~Plauschinn,
``Non-geometric backgrounds in string theory,''
Phys. Rept. \textbf{798} (2019), 1-122
[arXiv:1811.11203 [hep-th]].

\bibitem{Berman:2020tqn}
D.~S.~Berman and C.~D.~A.~Blair,
``The Geometry, Branes and Applications of Exceptional Field Theory,''
[arXiv:2006.09777 [hep-th]].



\bibitem{Gregory:1997te}
  R.~Gregory, J.~A.~Harvey and G.~W.~Moore,
  ``Unwinding strings and T-duality of Kaluza-Klein and H-monopoles,''
  Adv.\ Theor.\ Math.\ Phys.\  {\bf 1} (1997) 283
  [hep-th/9708086].

\bibitem{Tong:2002rq}
  D.~Tong,
  ``NS5-branes, T-duality and worldsheet instantons,''
  JHEP {\bf 0207} (2002) 013
  [hep-th/0204186].

\bibitem{Harvey:2005ab}
  J.~A.~Harvey and S.~Jensen,
  ``Worldsheet instanton corrections to the Kaluza-Klein monopole,''
  JHEP {\bf 0510} (2005) 028
  [hep-th/0507204].

\bibitem{Jensen:2011jna}
  S.~Jensen,
  ``The KK-monopole/NS5-brane in doubled geometry,''
  JHEP {\bf 1107} (2011) 088
  [arXiv:1106.1174 [hep-th]].

\bibitem{Kimura:2013fda}
  T.~Kimura and S.~Sasaki,
  ``Gauged linear sigma model for exotic five-brane,''
  Nucl.\ Phys.\ B {\bf 876} (2013) 493
  [arXiv:1304.4061 [hep-th]].

\bibitem{Kimura:2013zva}
  T.~Kimura and S.~Sasaki,
  ``Worldsheet instanton corrections to $5^2_2$-brane geometry,''
  JHEP {\bf 1308} (2013) 126
  [arXiv:1305.4439 [hep-th]].

\bibitem{Kimura:2018ain}
T.~Kimura, S.~Sasaki and K.~Shiozawa,
``Semi-doubled Gauged Linear Sigma Model for Five-branes of Codimension Two,''
JHEP \textbf{12} (2018), 095
[arXiv:1810.02169 [hep-th]].



\bibitem{Lust:2017jox}
  D.~L\"{u}st, E.~Plauschinn and V.~Vall Camell,
  ``Unwinding strings in semi-flatland,''
  JHEP {\bf 1707} (2017) 027
  [arXiv:1706.00835 [hep-th]].

\bibitem{Achmed-Zade:2018rfc}
I.~Achmed-Zade, M.~Hamilton, J.D., D.~L\"ust and S.~Massai,
``A note on T-folds and T$^{3}$ fibrations,''
JHEP \textbf{12} (2018), 020
[arXiv:1803.00550 [hep-th]].




\bibitem{Chatzistavrakidis:2013jqa}
  A.~Chatzistavrakidis, F.~F.~Gautason, G.~Moutsopoulos and M.~Zagermann,
  ``Effective actions of nongeometric five-branes,''
  Phys.\ Rev.\ D {\bf 89} (2014) no.6,  066004
  [arXiv:1309.2653 [hep-th]].

\bibitem{Kimura:2014upa}
  T.~Kimura, S.~Sasaki and M.~Yata,
  ``World-volume effective actions of exotic five-branes,''
  JHEP {\bf 1407} (2014) 127
  [arXiv:1404.5442 [hep-th]].

\bibitem{Kimura:2016anf}
  T.~Kimura, S.~Sasaki and M.~Yata,
  ``World-volume effective action of exotic five-brane in M-theory,''
  JHEP {\bf 1602} (2016) 168
  [arXiv:1601.05589 [hep-th]].

\bibitem{Sakatani:2016sko}
Y.~Sakatani and S.~Uehara,
``Branes in Extended Spacetime: Brane Worldvolume Theory Based on Duality Symmetry,''
Phys. Rev. Lett. \textbf{117} (2016) no.19, 191601
[arXiv:1607.04265 [hep-th]].

\bibitem{Blair:2017hhy}
  C.~D.~A.~Blair and E.~T.~Musaev,
  ``Five-brane actions in double field theory,''
  arXiv:1712.01739 [hep-th].

\bibitem{Sakatani:2020umt}
Y.~Sakatani and S.~Uehara,
``Born sigma model for branes in exceptional geometry,''
[arXiv:2004.09486 [hep-th]].



\bibitem{Hohm:2010pp}
  O.~Hohm, C.~Hull and B.~Zwiebach,
  ``Generalized metric formulation of double field theory,''
  JHEP {\bf 1008} (2010) 008
  [arXiv:1006.4823 [hep-th]].

\bibitem{Hohm:2011zr}
O.~Hohm, S.~K.~Kwak and B.~Zwiebach,
``Unification of Type II Strings and T-duality,''
Phys. Rev. Lett. \textbf{107} (2011), 171603
[arXiv:1106.5452 [hep-th]].

\bibitem{Hohm:2011dv}
O.~Hohm, S.~K.~Kwak and B.~Zwiebach,
``Double Field Theory of Type II Strings,''
JHEP \textbf{09} (2011), 013
[arXiv:1107.0008 [hep-th]].



\bibitem{Bergshoeff:2011se}
E.~A.~Bergshoeff, T.~Ortin and F.~Riccioni,
``Defect Branes,''
Nucl. Phys. B \textbf{856} (2012), 210-227
[arXiv:1109.4484 [hep-th]].




\bibitem{Adawi:1998ta}
T.~Adawi, M.~Cederwall, U.~Gran, B.~E.~W.~Nilsson and B.~Razaznejad,
``Goldstone tensor modes,''
JHEP \textbf{02} (1999), 001
[arXiv:hep-th/9811145 [hep-th]].



\bibitem{Brill:1964zz}
D.~R.~Brill and J.~B.~Hartle,
``Method of the Self-Consistent Field in General Relativity and its Application to the Gravitational Geon,''
Phys. Rev. \textbf{135} (1964), B271-B278.

\bibitem{Pope:1978zx}
C.~N.~Pope,
``Axial Vector Anomalies and the Index Theorem in Charged Schwarzschild and Taub - Nut Spaces,''
Nucl. Phys. B \textbf{141} (1978), 432-444



\bibitem{Park:2013mpa}
J.~H.~Park,
``Comments on double field theory and diffeomorphisms,''
JHEP \textbf{06} (2013), 098
[arXiv:1304.5946 [hep-th]].

\bibitem{Lee:2013hma}
K.~Lee and J.~H.~Park,
``Covariant action for a string in ''doubled yet gauged'' spacetime,''
Nucl. Phys. B \textbf{880} (2014), 134-154
[arXiv:1307.8377 [hep-th]].

\bibitem{Berman:2019izh}
D.~S.~Berman, C.~D.~A.~Blair and R.~Otsuki,
``Non-Riemannian geometry of M-theory,''
JHEP \textbf{07} (2019), 175
[arXiv:1902.01867 [hep-th]].




\bibitem{Mori:2019slw}
H.~Mori, S.~Sasaki and K.~Shiozawa,
``Doubled Aspects of Vaisman Algebroid and Gauge Symmetry in Double Field Theory,''
J. Math. Phys. \textbf{61} (2020) no.1, 013505
[arXiv:1901.04777 [hep-th]].

\bibitem{Ikeda:2020lxz}
N.~Ikeda and S.~Sasaki,
``Global Aspects of Doubled Geometry and Pre-rackoid,''
[arXiv:2006.08158 [math-ph]].

\bibitem{Mori:2020yih}
H.~Mori and S.~Sasaki,
``More on Doubled Aspects of Algebroids in Double Field Theory,''
[arXiv:2008.00402 [math-ph]].




\bibitem{Hohm:2011nu}
O.~Hohm and S.~K.~Kwak,
``N=1 Supersymmetric Double Field Theory,''
JHEP \textbf{03} (2012), 080
[arXiv:1111.7293 [hep-th]].

\bibitem{Jeon:2012hp}
I.~Jeon, K.~Lee, J.~H.~Park and Y.~Suh,
``Stringy Unification of Type IIA and IIB Supergravities under N=2 D=10 Supersymmetric Double Field Theory,''
Phys. Lett. B \textbf{723} (2013), 245-250
[arXiv:1210.5078 [hep-th]].



\bibitem{Blair:2014zba}
C.~D.~A.~Blair and E.~Malek,
``Geometry and fluxes of SL(5) exceptional field theory,''
JHEP \textbf{03} (2015), 144
[arXiv:1412.0635 [hep-th]].

\bibitem{Berman:2014hna}
  D.~S.~Berman and F.~J.~Rudolph,
  ``Strings, branes and the self-dual solutions of exceptional field theory,''
  JHEP {\bf 1505} (2015) 130
  [arXiv:1412.2768 [hep-th]].

\bibitem{Gunaydin:2016axc}
M.~Gunaydin, D.~Lust and E.~Malek,
``Non-associativity in non-geometric string and M-theory backgrounds, the algebra of octonions, and missing momentum modes,''
JHEP \textbf{11} (2016), 027
[arXiv:1607.06474 [hep-th]].

\bibitem{Lust:2017bwq}
D.~Lust, E.~Malek and M.~Syvari,
``Locally non-geometric fluxes and missing momenta in M-theory,''
JHEP \textbf{01} (2018), 050
[arXiv:1710.05919 [hep-th]].

\bibitem{Bakhmatov:2017les}
I.~Bakhmatov, D.~Berman, A.~Kleinschmidt, E.~Musaev and R.~Otsuki,
``Exotic branes in Exceptional Field Theory: the SL(5) duality group,''
JHEP \textbf{08} (2018), 021
[arXiv:1710.09740 [hep-th]].



\bibitem{Hohm:2013bwa}
O.~Hohm, D.~L\"ust and B.~Zwiebach,
``The Spacetime of Double Field Theory: Review, Remarks, and Outlook,''
Fortsch. Phys. \textbf{61} (2013), 926-966
[arXiv:1309.2977 [hep-th]].



\bibitem{Fukuma:1999jt}
M.~Fukuma, T.~Oota and H.~Tanaka,
``Comments on T dualities of Ramond-Ramond potentials on tori,''
Prog. Theor. Phys. \textbf{103} (2000), 425-446
[arXiv:hep-th/9907132 [hep-th]].



\end{thebibliography}
\end{document}